\documentclass[conference]{IEEEtran}
\IEEEoverridecommandlockouts
\usepackage{cite}
\usepackage[ruled,vlined]{algorithm2e}
\usepackage{graphicx}
\usepackage{subcaption}
\usepackage{amsmath,amsfonts}
\usepackage{algorithmic}
\usepackage{makecell}

\usepackage{array}

\usepackage{textcomp}
\usepackage{stfloats}
\usepackage{url}
\usepackage{xcolor}
\usepackage{verbatim}
\usepackage{multirow}
\usepackage{cite}
\usepackage{enumerate}
\usepackage{tabularx}
\usepackage{threeparttable}
\usepackage{amsmath}
\usepackage{booktabs}
\usepackage{amssymb}
\usepackage{pifont}
\usepackage{float}
\usepackage{multirow}
\usepackage{wrapfig}
\usepackage[colorlinks=true,linkcolor=blue,citecolor=blue,urlcolor=blue,breaklinks=true]{hyperref}
\usepackage{
	color,
	float,
	epsfig,
	graphics,
	graphicx,
	subcaption
}
\usepackage{xurl}
\usepackage[utf8]{inputenc}
\usepackage[compatibility=false]{caption}

\newenvironment{ditemize}{\begin{itemize}\setlength{\itemsep}{0pt}}{\end{itemize}}

\newcommand{\system}{\textsf{PrivQNN}}
\newcommand{\nop}[1]{}

\usepackage[most]{tcolorbox}

\newtcolorbox{takeaways}{
    colback=gray!5,
    colframe=black!40,
    boxrule=0.4pt,
    arc=1mm,
    left=5pt,
    right=5pt,
    top=4pt,
    bottom=4pt,
    before skip=10pt,
    after skip=10pt
}

\def\BibTeX{{\rm B\kern-.05em{\sc i\kern-.025em b}\kern-.08em
    T\kern-.1667em\lower.7ex\hbox{E}\kern-.125emX}}
\begin{document}

\title{Revisiting Privacy Risks in Quantized Deep Neural Networks}

\author{Jie Fu, Ren Li, Wendy Hui Wang \\Stevens Institute of Technology, Hoboken, USA\\jfu13@stevens.edu, rli56@stevens.edu, hwang4@stevens.edu}
\maketitle

\begin{abstract}
Model quantization has become a fundamental technique for enabling the efficient deployment of large-scale deep neural networks (DNNs) in data-intensive applications by reducing model size. Despite its widespread adoption, the impact of quantization on training data privacy ---- particularly vulnerability to {\em membership inference attacks} (MIAs) --- remains largely unexplored. In this paper, we present the first comprehensive empirical study of privacy risks in quantized DNNs, evaluating eight state-of-the-art MIA attacks across six representative quantization methods. Our results reveal that practical quantization levels (4-8 bits) have limited impact on membership privacy risks, with quantized models remaining largely as vulnerable as their full-precision counterparts. Although aggressive low-bit quantization (e.g., 2-bit) can provide meaningful privacy gains, these benefits come at the cost of substantial utility degradation. To understand the underlying causes of these observations, we systematically analyze how different quantization strategies affect model behaviors and outputs, and how these changes influence privacy vulnerability.
Beyond evaluating attack risks, we further investigate whether existing privacy defenses developed for full-precision DNNs can effectively protect QNNs. Our results show that these defenses provide only limited additional privacy protection when applied to quantized models, while further amplifying utility degradation. Our study provides practical guidance for privacy-aware deployment of quantized DNNs and motivates the development of new attack and defense mechanisms for QNNs. 
Our code and datasets are publicly available at \url{https://github.com/JeffffffFu/Priv-QDNN}.

\end{abstract}

\section{Introduction}\label{sec-intro}
Deep Neural Networks (DNNs) have achieved remarkable success in numerous practical applications. 
However, deploying high-performance DNNs often incurs substantial computational and memory costs, which limit their applicability in real-world settings, particularly on resource-constrained devices such as mobile phones and IoT platforms. To address these challenges, model quantization has emerged as a widely adopted technique that reduces model size and accelerates inference by representing weights and activations using lower-precision numerical formats (e.g., 8 bits), while maintaining acceptable accuracy \cite{rokh2023comprehensive,gholami2022survey}.

Despite these advantages, the privacy implications of model quantization remain largely unexplored. Existing studies on privacy attacks have primarily focused on full-precision DNNs \cite{aubinais2025membership,shokri2017membership,salem2019ml,nasr2018machine,duddu2022inferring,he2019model,song2021systematic}, leaving a critical gap in understanding how quantization influences the privacy vulnerability of neural networks. Intuitively, by reducing the precision of model parameters, quantization alters both a model's output distributions and its internal behaviors, including overfitting and memorization. Since these properties are key factors exploited by privacy attacks, particularly {\em membership inference attacks} (MIAs) \cite{shokri2017membership,carlini2022membership}, quantization has the potential to alter membership privacy risks. This intuition has motivated prior work to explore quantization as a potential means of mitigating MIAs \cite{kowalski2022towards}.

However, recent evidence suggests that the relationship between quantization and privacy is far from straightforward. Li et al.~\cite{li2025compleak} reported that, for certain models, existing MIAs achieve nearly identical performance on 8-bit quantized models and their full-precision counterparts, indicating that practical quantization may have little impact on membership privacy. These seemingly conflicting observations raise a fundamental question: {\em How does model quantization affect membership privacy risks, and at what level of compression do meaningful privacy changes emerge?}

Recent studies have investigated the impact of model compression on various trustworthiness aspects of machine learning models such as  fairness, toxicity, bias, robustness, and truthfulness~\cite{hong2024decoding,xu2024beyond,fu2025quantized}. However, privacy has received little attention in this line of research. While several recent works have evaluated MIAs against quantized models~\cite{kowalski2022towards,famili2023deep,jin2023membership,aubinais2025membership,zhang2025bits,li2025compleak}, they consider only a limited set of quantization methods, attack techniques, or experimental settings. As a result, a comprehensive understanding of how different quantization strategies affect membership privacy across diverse attack scenarios remains lacking.

\begin{table}[t]
\centering
\caption{Comparison between existing studies evaluating membership inference
attacks on quantized models and ours. }
\label{tab:related_quant_mia}
\scalebox{0.98}{
\begin{tabular}{l|c|c|c|c|c}
\hline
\textbf{Work}
& \textbf{PTQ}
& \textbf{QAT}
& \textbf{Precision}
& \textbf{Threat}
& \textbf{Granularity} \\
\hline

\cite{kowalski2022towards}
& \checkmark
& --
& Fixed
& Black-box
& Population-level \\\hline

\cite{aubinais2025membership,zhang2025bits}
& \checkmark
& --
& Fixed
& Black-box
& Example-level  \\\hline

\multirow{2}{*}{\cite{li2025compleak}}
& \multirow{2}{*}{--}
& \multirow{2}{*}{\checkmark}
& \multirow{2}{*}{Fixed}
& \multirow{2}{*}{Black-box}
& Population \& \\

& 
& 
& 
& 
& Example\\\hline

\multirow{2}{*}{\textbf{Ours}}
& \multirow{2}{*}{\checkmark}
& \multirow{2}{*}{\checkmark}
& Fixed \&  
& Black- \&
& Population \&  \\

& 
& 
& Mixed
& White-box
& Example\\\hline

\end{tabular}
}
\end{table}

In this work, we present the first comprehensive empirical study of membership privacy risks in quantized neural networks (QNNs). We systematically evaluate eight representative MIAs across six state-of-the-art quantization techniques. The attacks span both black-box and white-box threat models and operate at both the population and per-example levels, while the evaluated quantization methods cover a broad spectrum of design choices, including post-training quantization (PTQ) and quantization-aware training (QAT), as well as fixed-precision and mixed-precision quantization. The difference between the existing works and ours is summarized in Table~\ref{tab:related_quant_mia}. 
Our key contributions are summarized as follows:

\begin{itemize}

\item  We rigorously assess a broad range of QNNs, providing insights for the design of efficient DNNs with stronger resilience against MIAs.

    \item Our empirical study reveals  that quantization should not be considered a privacy-preserving mechanism: practical compression levels commonly used in deployment (4-8 bits) largely preserve membership privacy risks, while meaningful privacy improvements require extreme low-bit quantization that severely compromises model utility.

    \item To understand the underlying causes of quantization's impact on privacy, we analyze its effects on key model behaviors associated with membership leakage, including model overfitting and output distribution. Our results demonstrate that the impact of quantization varies across  model architectures and attack settings.

    \item We further evaluate the effectiveness of existing MIA defense mechanisms for full-precision DNNs, including {\em differential privacy}~\cite{dwork2006differential,abadi2016deep}, {\em dropout}~\cite{kaya2020effectiveness,yin2021defending}, and {\em adversarial learning}~\cite{nasr2018machine,li2021deepobfuscator,wang2021improving}, when applied to QNNs. Our results show that these defenses provide only limited additional privacy protection, while their perturbations can further amplify the utility degradation introduced by quantization.
\end{itemize}

Our findings call for a new paradigm of privacy-aware model compression, where quantization, privacy protection, and model utility are jointly optimized to enable secure and efficient deployment of DNNs in real-world applications. 
\section{Background and Related Work}\label{sec-related}

\subsection{Quantized Deep Neural Network}

Quantization is a fundamental technique for reducing the size of DNNs and accelerating inference. It converts the parameters and/or intermediate activations (typically 32-bit floating-point format) to lower bit-width representations. 
A commonly used quantization function is defined as: 
$q = round(\frac{x}{\Delta}) + z$,    
where $\Delta$ is the quantization step size decided by the bit-width $b$: $\Delta =\frac{x_{\text{max}} - x_{\text{min}}}{2^b-1}$, $z$ is the {\em zero-point} that represents the real value $0$ in the quantized space, and $round(\cdot)$ maps the scaled input to the nearest integer. In this work, we focus primarily on the configuration of the bit-width $b$ value, which is the dominant factor controlling the level of quantization noise and is the most  widely tuned parameter in practice, 
while fixing the remaining quantization parameters across all experiments. This design choice allows us to isolate the impact of precision on privacy leakage without introducing additional confounding factors. 

\smallskip
{\bf PTQ vs. QAT.} Existing quantization mechanisms for DNNs can be grouped into two types~\cite{rokh2023comprehensive}: \textit{Post-Training Quantization (PTQ)} and \textit{Quantization-Aware Training (QAT)}. PTQ performs quantization after training a model with full-precision weights and activations, while QAT incorporates quantization effects during training, allowing the model to adapt to quantization noise. In this study, we examine PTQ and QAT quantization methods. %

\smallskip
{\bf Fixed-precision vs. Mixed-precision.}  In fixed-precision quantization, weights and activations are reduced to the same bit-width uniformly. In contrast, mixed-precision quantization approaches assigns different bit-widths to the weights at different layers based on their sensitivity to quantization. In this study, we examine both fixed-precision and mixed-precision quantization methods. 

\subsection{Membership Inference Attacks against DNNs}
{\em Membership Inference Attack} (MIA) \cite{aubinais2025membership,shokri2017membership,salem2019ml,nasr2018machine} is one of the most widely-studied privacy attacks against machine learning models in the literature. Given a trained model, it aims to determine whether a data sample was present in the training dataset of the model.

Based on the adversary's access to the victim model, existing MIAs can be categorized into black-box and white-box settings.

\smallskip
{\bf Black-box MIA.} A black-box adversary only can access model outputs (e.g., confidence scores) to distinguish members and non-members \cite{shokri2017membership,salem2019ml,song2021systematic}. 
Based on the granularity of the attack, existing black-box MIAs can be classified into {\em per-example} and {\em population-based} ones~\cite{hayes2024inexact}. A population-based MIA trains an attack model as a binary classifier using outputs of samples drawn from a data distribution~\cite{shokri2017membership,salem2019ml}. 
It typically uses shadow datasets to train shadow models, and then uses member/non-member outputs from those shadow models to train the attack classifier. In contrast, a per-example MIA trains multiple shadow models with and without a specific target sample and measures how prediction confidence changes across these models~\cite{zarifzadeh2024low}. It then computes a membership likelihood ratio for that target sample. 

\smallskip
{\bf White-box MIA.} 
A white-box adversary has access to internal model states such as parameters, gradients, and intermediate representations \cite{leino2020stolen,sablayrolles2019white}. Such access enables the attacker to exploit fine-grained signals such as per-sample gradients, gradient norms, and hidden-layer activations  to distinguish members from non-members  \cite{nasr2019comprehensive}. 

In this study, we consider both white-box and black-box MIAs at per-example and population-based granularity. More details of our attacks can be found in Section \ref{sc:quantmodel}.

\subsection{Membership Privacy of Quantized DNNs}
A few recent studies have begun to examine membership privacy risks in quantized models. Kowalski et al.~\cite{kowalski2022towards} conducted the first empirical investigation into how model quantization affects the resistance of DNNs to MIAs. However, their analysis is limited to a single quantization method, DoReFa-Net~\cite{zhou2016dorefa}. Aubinais et al.~\cite{aubinais2025membership} extended this line of work by providing a more rigorous theoretical analysis of how quantization influences membership inference vulnerability, but both their theory and experiments are restricted to simplified quantization schemes modeled as linear functions. Zhang et al.~\cite{zhang2025bits} studied the privacy--utility trade-off of three PTQ methods at 4-bit, 2-bit, and 1.58-bit precision. However, their evaluation is limited to fixed-precision PTQ configurations and per-example black-box MIAs. Li et al.~\cite{li2025compleak} investigated membership leakage induced by model compression techniques, including pruning, quantization, and weight clustering. Their results show that existing MIAs achieve nearly identical performance on 8-bit quantized models and their full-precision counterparts, suggesting that widely used quantization may not provide meaningful membership privacy protection. However, their study considers only the quantization methods implemented in PyTorch Mobile and evaluates only 8-bit quantized models.
Several recent studies have also proposed MIAs specifically tailored to quantized models~\cite{jin2023membership,famili2023deep,li2025compleak}. While these attacks demonstrate that quantization introduces new attack opportunities, they are designed for specific quantization schemes and attack settings. Consequently, it remains unclear how quantization affects membership privacy more broadly across diverse quantization methods, precision levels, and attack models, including the existing MIAs that constitute the standard threat model for full-precision DNNs.

\section{Our Evaluation Framework}

Understanding the impact of quantization on a model's vulnerability to MIAs requires a comprehensive empirical evaluation. Thus, we conduct a comprehensive empirical study spanning a diverse set of datasets, DNN architectures, quantization methods, and MIA techniques.

\subsection{Datasets and DNNs} 

Following prior pioneering studies \cite{salem2019ml,shokri2017membership,zarifzadeh2024low}, we consider four benchmark datasets: two image datasets, \textit{CIFAR-100}~\cite{krizhevsky2009learning} and \textit{CINIC-10}~\cite{darlow2018cinic}, and two tabular datasets, \textit{Location} and \textit{Texas}. For the two image datasets, we train two widely adopted DNN architectures, \textit{ResNet-18}~\cite{he2016deep} and \textit{MobileNetV2}~\cite{sandler2018mobilenetv2}, which are representative models used in practical deployments. For \textit{Location}
and \textit{Texas} 
 datasets, we train a five-layer Multi-Layer Perceptron (MLP) model. 
 More details of the models and the datasets can be found in Appendix~\ref{sec:appendix_dataset_stats}.

\subsection{Quantization Methods}
\label{sc:quantmodel}

\begin{table}[!t]
\centering
\caption{Categorization of the evaluated quantization methods}
\begin{tabular}{c|c|c}
\hline
{\bf Method} & {\bf PTQ or QAT} & {\bf Precision type} \\
\hline
PQ \cite{torchao} & PTQ & Fixed  \\\hline
AdaRound \cite{nagel2020up} & PTQ & Fixed\\
\hline
QDrop \cite{wei2022qdrop} & PTQ & Fixed\\
\hline
PACT \cite{choi2018pact} & QAT  & Fixed\\
\hline
BSQ \cite{yang2021bsq} & QAT & Mixed \\
\hline
HAWQ \cite{dong2019hawq} & QAT & Mixed \\
\hline
\end{tabular}
\label{tab:quant}
\end{table}

We deploy six state-of-the-art quantization methods that are widely adopted for DNNs in practice. These methods span several major categories of quantization techniques. Their taxonomy is summarized in Table~\ref{tab:quant}. 
\ding{172} \textbf{PyTorch-Quant (PQ)} \cite{torchao}
 is a native quantization library integrated with PyTorch. We adopt per-tensor quantization with a uniform quantization scheme as the mapping function. 
\ding{173} \textbf{AdaRound} \cite{nagel2020up}
is a PTQ quantization mechanism that adaptively determines whether to round the weights to their nearest values by considering the characteristics of both the input data and the task loss. 
\ding{174} \textbf{QDrop} \cite{wei2022qdrop}
incorporates activation quantization during output reconstruction. 
\ding{175} \textbf{PACT} \cite{choi2018pact}
 automatically optimizes the quantization scales during model training. 
\ding{176} \textbf{BSQ} \cite{yang2021bsq} is a mixed-precision method that 
incorporates a sparsity-inducing regularizer with the objective function of the model to optimize each bit of quantized weights during model training. 
\ding{177}\textbf{HAWQ} \cite{dong2019hawq}
is a mixed-precision quantization technique by which higher precision is used for certain “sensitive” layers of the network, and lower precision for “non-sensitive” layers. 

In our empirical evaluation, we apply quantization to both weights and activations across all methods, considering the bit-widths $b \in \{2, 4, 6, 8\}$. We use the same bit-width for both weights and activation.

\subsection{Membership Inference Attacks}  
\label{sc:mia}

\begin{figure}[t!]
    \centering
    \includegraphics[width=\linewidth]{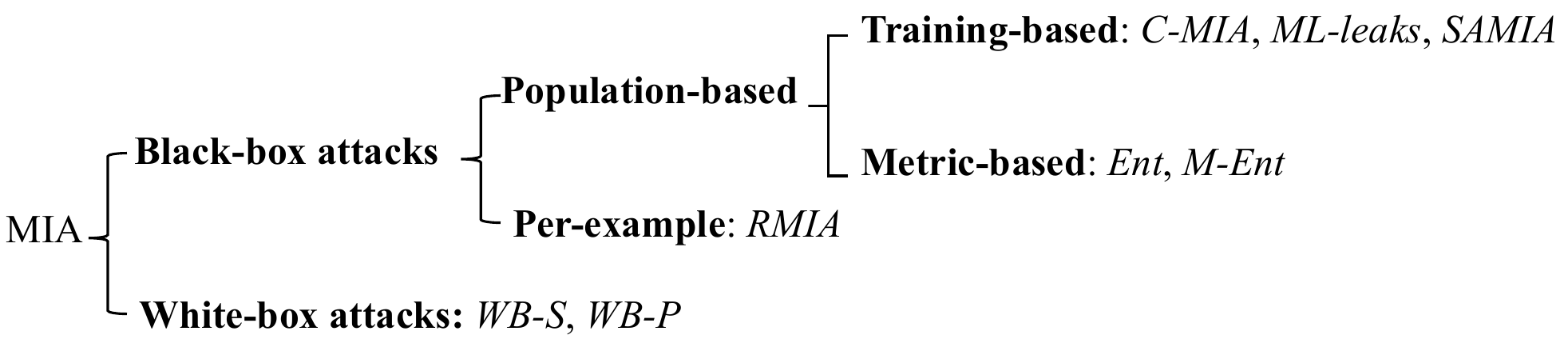}
    \caption{A taxonomy of the MIA attacks in the study}
    \label{fig:MIA-attacks}
\end{figure}

We evaluate eight state-of-the-art MIAs against DNNs, including six for the black-box setting and two for the white-box setting. The taxonomy of these MIAs is shown in Figure~\ref{fig:MIA-attacks}.

\subsubsection{Black-box Setting}
We employ six black-box MIAs, five of which are at the population granularity, and one is at the per-example granularity. 

{\bf Population-level granularity.} We conduct five population-level MIAs, three of which are training-based and two are metric-based.

\begin{itemize}
\item \textbf{Training-based attacks.} These attacks train one or more shadow models and leverage the outputs of the trained shadow models on member and non-member samples to train a meta-classifier for membership inference. In this study, we employ three state-of-the-art training-based MIAs: \textbf{C-MIA}~\cite{nasr2018machine}
     derives the attack feature from the posterior output of the ground-truth label; \textbf{ML-Leaks}~\cite{salem2019ml}      %
   derives the attack features from the sorted high-confidence posterior values;  \textbf{SAMIA}~\cite{yuan2022mia_pruning}   %
   derives its attack features from prediction confidence, prediction sensitivity, and the ground-truth label.  Following prior work \cite{chen2021machine}, we train a three-layer fully connected Multilayer Perceptron (MLP) as the meta-classifier. The three layers consist of 32, 16, and 2 neurons. 
   
\item \textbf{Metric-based attacks.} These attacks use a threshold  to infer membership based on the metric values computed from the model outputs ~\cite{song2021systematic,yeom2018privacy,salem2019ml}. 
In our work, we consider two metrics:  \textbf{Entropy} (Ent) ~\cite{salem2019ml}, which 
     uses the predictive entropy $ H(x) = - \sum_i p_i \log p_i.$ as the membership score and selects a threshold of the entropy to distinguish members from non-members, and \textbf{M-Entropy (M-Ent)}~\cite{song2021systematic}
     which 
   uses the label-aware modified entropy, which incorporates the predictive entropy with the ground-truth label and thus is more sensitive than standard entropy for membership inference. 
\end{itemize}

\smallskip\textbf{Example-level granularity.} In the study, we employ \textbf{RMIA}~\cite{zarifzadeh2024low}, 
a low-cost and high-power variant of LiRA \cite{carlini2022membership}.  
RMIA is primarily based on the pairwise likelihood ratio test that compares the victim model's output on a target sample against its outputs on random population samples. A sample is inferred as a member if it
dominates a sufficiently large fraction of population samples
under this test.

\subsubsection{White-box Adversary.}
We consider two white-box MIAs~\cite{nasr2019comprehensive}: 
\textbf{WB-S}~\cite{nasr2019comprehensive} whose attack features are derived from the softmax probability on the true class and the $\ell_2$ norm of the gradient on the last linear layer's weights,  %
 and 
\textbf{WB-P}~\cite{nasr2019comprehensive}, which uses the same white-box features as WB-S, but 
 avoids training a full shadow model and instead uses an auxiliary dataset to derive the features. We randomly sample an auxiliary dataset from the same distribution as the target model's training data, with a size equal to 20\% of the target training set.

\subsection{Evaluation Metrics} 
\label{sc:metrics}
We consider two  metrics widely adopted in the MIA literature~\cite{salem2019ml,nasr2019comprehensive,zarifzadeh2024low} for attack evaluation:  
\begin{ditemize}
\item \textbf{Attack accuracy.} It measures the proportion of samples that are correctly classified as either members or non-members.

\item \textbf{TPR @ low FPR} measures the true positive rate (TPR) (i.e., the fraction of correctly identified member samples) at a predefined low false positive rate (FPR) threshold (e.g., 1\%). 
\end{ditemize}

In addition, we report {\em model utility} as the proportion of samples in the training dataset that are correctly classified by the model. 

\section{Impact of Quantization on MIA Effectiveness}
\label{sec-evaluation}

In this work, we seek to answer the following research question through an extensive empirical study: {\em Does quantization affect the privacy risks of DNNs against MIAs?} All experiments are conducted on a server equipped with an NVIDIA A100 GPU and 40 GB of memory. 
All experimental results are averaged over five trials, with both the mean and standard deviation reported.   
Due to space constraints, the results in the main text are the attack accuracy results. 
The TPR@1\% FPR results can be found in Appendix~\ref{sec:appendix_mia_acc_cinic10}.   

\begin{table*}[!t]
    \centering
    \caption{Attack accuracy (\%) across 4-bit quantized models in different settings.}
    \label{tab:MIA-attack-4bit}
    \scalebox{1}{
    \begin{tabular}{c|c|c|c|c|c|c|c|c|c}
    \hline
    \multirow{2}{*}{\textbf{Dataset\&Model}} & \multirow{2}{*}{\textbf{Quantization}}
    & \multicolumn{6}{c|}{\textbf{Black-box MIA}} 
    & \multicolumn{2}{c}{\textbf{White-box MIA}} \\
    \cline{3-10}
    & & \textbf{C-MIA} & \textbf{ML-leaks} & \textbf{SAMIA} & \textbf{Ent} & \textbf{M-Ent} & \textbf{RMIA} & \textbf{WB-S} & \textbf{WB-P} \\
    \hline
    
    \multirow{7}{*}{\makecell{CIFAR-100 \\ ResNet-18}}
    & Full-precision
    & 85.2$_{\pm0.4}$ & 68.1$_{\pm1.2}$ & 87.1$_{\pm0.4}$ & 77.5$_{\pm1.1}$ & 85.2$_{\pm0.4}$ & 87.1$_{\pm0.2}$ & 87.7$_{\pm0.5}$ & 89.0$_{\pm0.9}$ \\\cline{2-10}
    & PQ
    & 79.4$_{\pm0.9}$ & 62.4$_{\pm0.6}$ & 81.7$_{\pm0.8}$ & 66.8$_{\pm1.4}$ & 79.3$_{\pm1.0}$ & 82.5$_{\pm0.8}$ & 83.0$_{\pm1.0}$ & 84.3$_{\pm0.8}$ \\
    & AdaRound
    & 80.6$_{\pm0.4}$ & 60.5$_{\pm1.4}$ & 81.9$_{\pm0.3}$ & 66.8$_{\pm1.0}$ & 80.6$_{\pm0.5}$ & 82.6$_{\pm0.3}$ & 83.2$_{\pm0.4}$ & 83.9$_{\pm0.6}$ \\
    & QDrop
    & 80.0$_{\pm0.5}$ & 62.6$_{\pm0.1}$ & 82.1$_{\pm0.6}$ & 65.4$_{\pm0.5}$ & 79.8$_{\pm0.5}$ & 81.4$_{\pm0.6}$ & 83.9$_{\pm0.5}$ & 85.7$_{\pm0.2}$ \\
    & HAWQ
    & 84.2$_{\pm0.4}$ & 67.5$_{\pm5.4}$ & 85.6$_{\pm0.9}$ & 76.1$_{\pm1.9}$ & 84.4$_{\pm0.4}$ & 87.0$_{\pm0.4}$ & 87.3$_{\pm1.1}$ & 88.5$_{\pm1.1}$ \\
    & PACT
    & 85.1$_{\pm0.2}$ & 68.0$_{\pm0.8}$ & 86.3$_{\pm0.1}$ & 76.3$_{\pm0.7}$ & 84.3$_{\pm0.3}$ & 86.0$_{\pm0.5}$ & 87.0$_{\pm0.5}$ & 88.3$_{\pm0.3}$ \\
    & BSQ
    & 67.4$_{\pm0.6}$ & 46.5$_{\pm0.5}$ & 69.3$_{\pm1.2}$ & 51.9$_{\pm0.3}$ & 67.2$_{\pm0.9}$ & 70.2$_{\pm0.5}$ & 72.1$_{\pm1.3}$ & 78.5$_{\pm1.9}$ \\
    \hline
    
    \multirow{7}{*}{\makecell{CINIC-10 \\ MobileNetV2}}
    & Full-precision
    & 77.0$_{\pm0.5}$ & 71.4$_{\pm1.0}$ & 77.8$_{\pm0.3}$ & 64.6$_{\pm0.3}$ & 77.1$_{\pm0.6}$ & 77.8$_{\pm0.4}$ & 78.1$_{\pm0.1}$ & 79.0$_{\pm0.3}$ \\\cline{2-10}
    & PQ
    & 53.3$_{\pm0.7}$ & 50.1$_{\pm0.3}$ & 53.3$_{\pm0.6}$ & 50.0$_{\pm0.4}$ & 53.2$_{\pm0.9}$ & 53.4$_{\pm0.5}$ & 54.7$_{\pm0.8}$ & 55.0$_{\pm0.4}$ \\
    & AdaRound
    & 56.2$_{\pm0.8}$ & 53.4$_{\pm0.6}$ & 58.5$_{\pm0.9}$ & 51.0$_{\pm0.5}$ & 56.2$_{\pm1.1}$ & 59.2$_{\pm0.7}$ & 61.9$_{\pm1.2}$ & 62.2$_{\pm0.6}$ \\
    & QDrop
    & 55.1$_{\pm0.5}$ & 52.0$_{\pm0.4}$ & 56.6$_{\pm0.7}$ & 50.4$_{\pm0.5}$ & 55.1$_{\pm0.8}$ & 56.8$_{\pm0.6}$ & 58.3$_{\pm0.9}$ & 58.5$_{\pm0.5}$ \\
    & HAWQ
    & 58.6$_{\pm1.2}$ & 57.3$_{\pm1.4}$ & 56.8$_{\pm0.9}$ & 53.1$_{\pm0.8}$ & 58.9$_{\pm1.0}$ & 57.6$_{\pm0.7}$ & 59.4$_{\pm1.1}$ & 59.7$_{\pm0.8}$ \\
    & PACT
    & 59.2$_{\pm0.9}$ & 56.6$_{\pm0.8}$ & 59.7$_{\pm1.0}$ & 52.0$_{\pm0.6}$ & 59.4$_{\pm1.3}$ & 59.9$_{\pm0.3}$ & 60.1$_{\pm0.7}$ & 60.3$_{\pm1.2}$ \\
    & BSQ
    & 54.9$_{\pm1.1}$ & 55.8$_{\pm0.9}$ & 65.5$_{\pm1.2}$ & 51.7$_{\pm0.7}$ & 50.6$_{\pm1.4}$ & 65.8$_{\pm0.5}$ & 67.4$_{\pm0.8}$ & 67.7$_{\pm1.3}$ \\
    \hline
    
    \multirow{7}{*}{\makecell{Texas\\ MLP}}
    & Full-precision
    & 71.3$_{\pm2.8}$ & 57.0$_{\pm1.6}$ & 72.9$_{\pm2.4}$ & 62.1$_{\pm2.8}$ & 71.4$_{\pm3.1}$ & 71.9$_{\pm1.4}$ & 78.1$_{\pm2.9}$ & 78.9$_{\pm2.0}$ \\\cline{2-10}
    & PQ
    & 70.4$_{\pm0.8}$ & 56.0$_{\pm0.7}$ & 71.2$_{\pm0.5}$ & 59.4$_{\pm0.9}$ & 70.1$_{\pm0.8}$ & 71.4$_{\pm0.2}$ & 75.0$_{\pm0.5}$ & 75.4$_{\pm0.5}$ \\
    & AdaRound
    & 70.3$_{\pm1.3}$ & 55.3$_{\pm0.6}$ & 71.7$_{\pm1.2}$ & 59.1$_{\pm1.1}$ & 70.1$_{\pm1.4}$ & 70.6$_{\pm0.6}$ & 74.3$_{\pm1.4}$ & 74.8$_{\pm1.2}$ \\
    & QDrop
    & 70.7$_{\pm0.1}$ & 56.6$_{\pm0.0}$ & 71.7$_{\pm0.1}$ & 60.8$_{\pm0.0}$ & 70.9$_{\pm0.1}$ & 71.9$_{\pm0.3}$ & 74.0$_{\pm0.5}$ & 75.9$_{\pm0.2}$ \\
    & HAWQ
    & 70.5$_{\pm0.2}$ & 56.1$_{\pm0.2}$ & 70.2$_{\pm2.2}$ & 61.2$_{\pm0.2}$ & 70.7$_{\pm0.2}$ & 71.8$_{\pm0.5}$ & 77.4$_{\pm2.8}$ & 78.0$_{\pm0.5}$ \\
    & PACT
    & 70.6$_{\pm0.6}$ & 55.2$_{\pm0.2}$ & 70.5$_{\pm0.3}$ & 60.0$_{\pm0.6}$ & 71.1$_{\pm0.6}$ & 71.5$_{\pm0.2}$ & 77.3$_{\pm0.1}$ & 78.0$_{\pm0.3}$ \\
    & BSQ
    & 68.0$_{\pm0.2}$ & 56.5$_{\pm0.4}$ & 70.0$_{\pm0.2}$ & 58.1$_{\pm0.1}$ & 68.2$_{\pm0.2}$ & 71.1$_{\pm0.6}$ & 73.6$_{\pm0.1}$ & 74.1$_{\pm0.2}$ \\
    \hline
    
    \multirow{7}{*}{\makecell{Location \\ MLP} }
    & Full-precision
    & 73.8$_{\pm0.2}$ & 53.8$_{\pm0.3}$ & 74.0$_{\pm0.4}$ & 54.7$_{\pm0.8}$ & 73.1$_{\pm0.2}$ & 74.3$_{\pm0.7}$ & 79.8$_{\pm0.4}$ & 80.8$_{\pm0.7}$ \\\cline{2-10}
    & PQ
    & 72.4$_{\pm0.5}$ & 52.7$_{\pm0.7}$ & 72.1$_{\pm0.2}$ & 54.5$_{\pm0.2}$ & 72.1$_{\pm0.4}$ & 74.4$_{\pm1.2}$ & 76.5$_{\pm0.1}$ & 77.4$_{\pm0.6}$ \\
    & AdaRound
    & 73.4$_{\pm0.9}$ & 52.0$_{\pm0.1}$ & 71.8$_{\pm0.4}$ & 53.7$_{\pm0.4}$ & 73.4$_{\pm0.9}$ & 73.8$_{\pm0.5}$ & 75.4$_{\pm0.0}$ & 76.8$_{\pm1.1}$ \\
    & QDrop
    & 71.4$_{\pm0.7}$ & 51.6$_{\pm1.5}$ & 66.5$_{\pm0.9}$ & 49.5$_{\pm0.8}$ & 71.5$_{\pm0.5}$ & 71.3$_{\pm0.2}$ & 73.7$_{\pm0.1}$ & 75.0$_{\pm0.7}$ \\
    & HAWQ
    & 70.1$_{\pm0.5}$ & 53.3$_{\pm1.4}$ & 62.6$_{\pm0.3}$ & 51.5$_{\pm0.1}$ & 70.4$_{\pm0.4}$ & 73.4$_{\pm0.1}$ & 74.6$_{\pm0.5}$ & 75.7$_{\pm0.6}$ \\
    & PACT
    & 72.4$_{\pm1.6}$ & 52.2$_{\pm1.7}$ & 73.2$_{\pm0.5}$ & 54.2$_{\pm0.8}$ & 72.2$_{\pm1.4}$ & 73.4$_{\pm0.4}$ & 78.9$_{\pm0.8}$ & 79.7$_{\pm1.3}$ \\
    & BSQ
    & 68.2$_{\pm0.8}$ & 51.2$_{\pm0.5}$ & 64.2$_{\pm0.8}$ & 50.3$_{\pm0.1}$ & 68.1$_{\pm0.8}$ & 66.8$_{\pm0.3}$ & 70.2$_{\pm0.2}$ & 71.6$_{\pm0.2}$ \\
    \hline
    \end{tabular}
    }
    \end{table*}

\subsection{MIA Effectiveness}

Since 4-bit and 8-bit are the most widely adopted quantization configurations for DNN deployment~\cite{yao2021hawq,dong2022finding}, and prior work has shown that 8-bit quantization yields nearly identical MIA performance to full-precision models~\cite{li2025compleak}, we first investigate whether 4-bit quantization can alter membership privacy risks. As reported in Table~\ref{tab:MIA-attack-4bit}, the impact of 4-bit quantization is highly dependent on the quantization method, attack, and target model. In most settings, quantized models remain nearly as vulnerable to MIAs as their full-precision counterparts. The only notable exception is the MobileNetV2 model trained on the CINIC-10 dataset, where 4-bit quantization substantially reduces MIA effectiveness, with attack accuracy approaching random guessing in most cases. We investigate the underlying factors contributing to this behavior in Section~\ref{sc:effect}.

\begin{figure}[!t]
\centering
\includegraphics[width=0.45\textwidth]{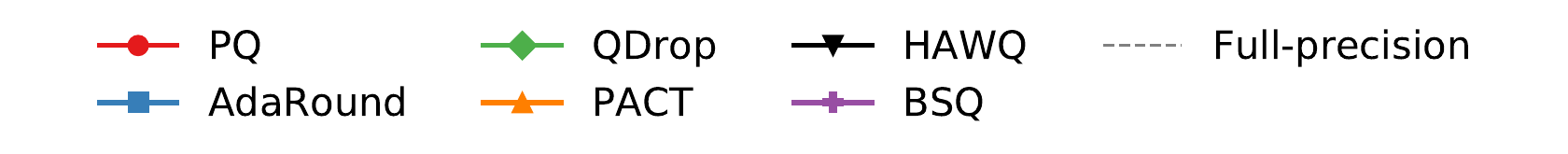}

\begin{subfigure}[b]{0.24\textwidth}
    \centering
    \includegraphics[width=\linewidth,height=0.72\linewidth,keepaspectratio]{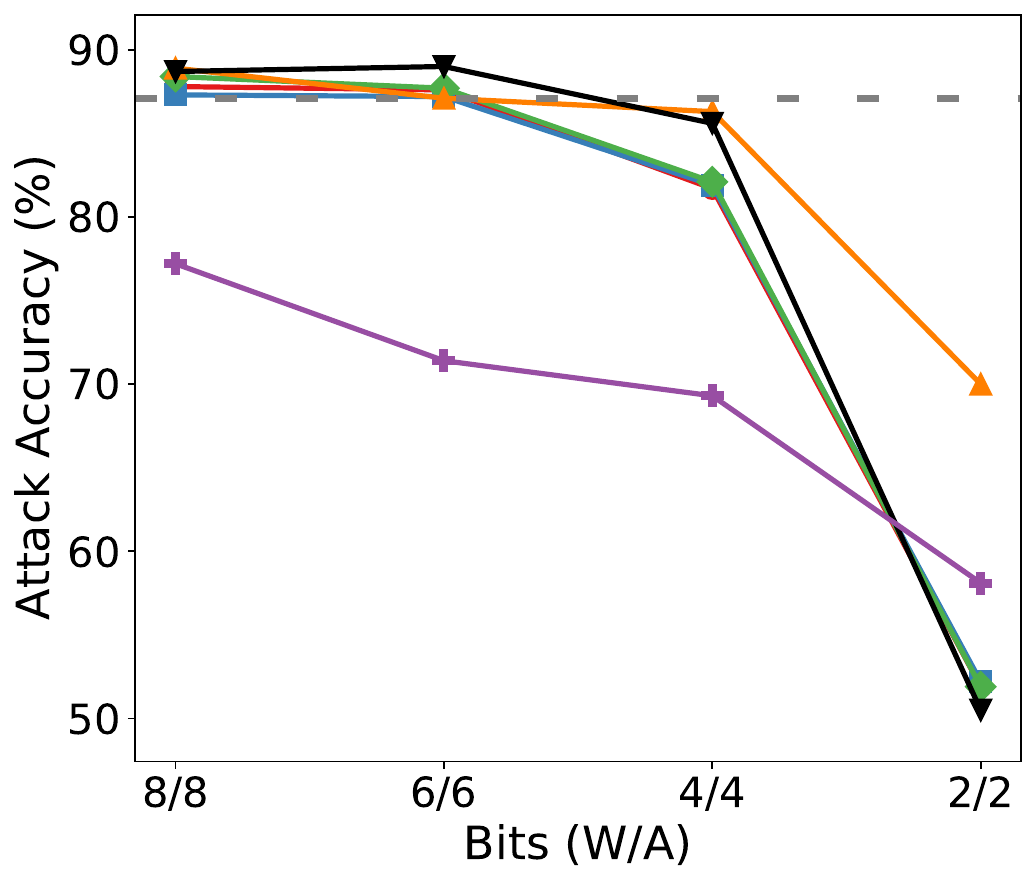}
    \caption{SAMIA attack}
\end{subfigure}
\hfill
\begin{subfigure}[b]{0.24\textwidth}
    \centering
    \includegraphics[width=\linewidth,height=0.72\linewidth,keepaspectratio]{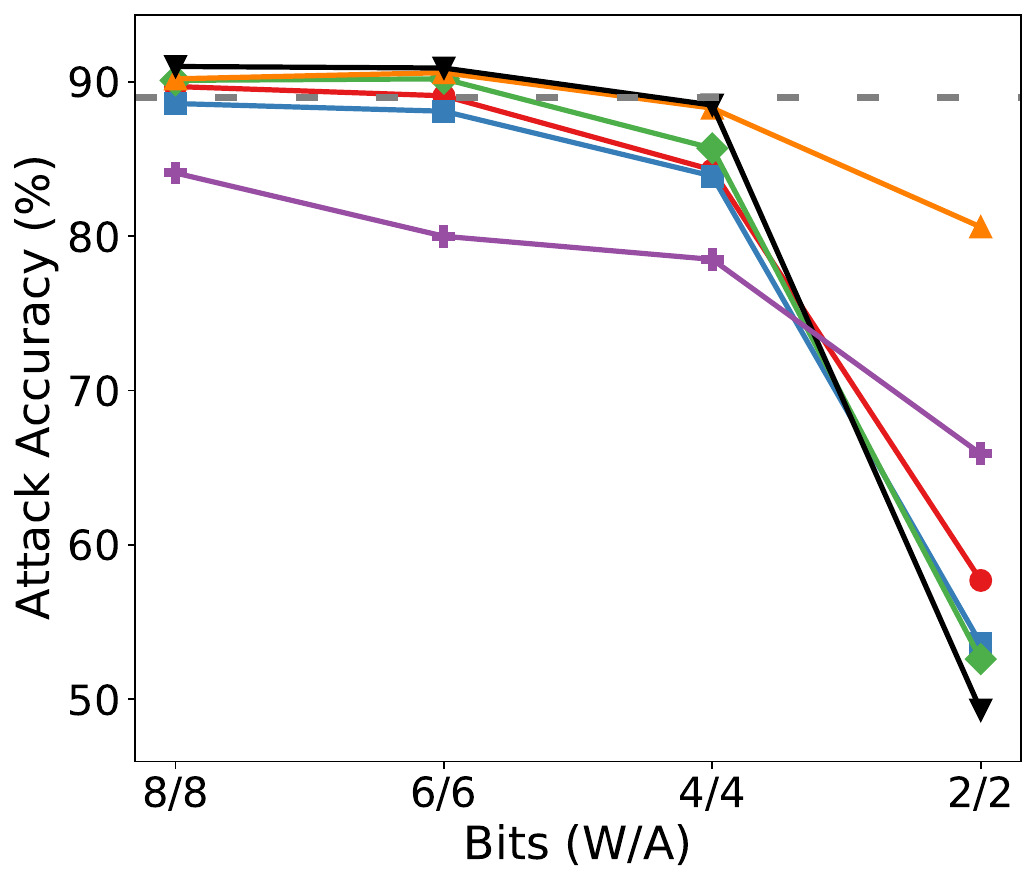}
    \caption{WB-P attack}
\end{subfigure}

\caption{Attack accuracy under various quantization levels (ResNet-18, CIFAR-100 dataset). }
\label{fig:mia_attack_acc_cifar100_by_bits}
\end{figure}

Since 4-bit quantization has been demonstrated to fail to provide meaningful privacy protection in many settings, we next examine whether more aggressive quantization can mitigate membership privacy risks. To isolate the effect of quantization level, we focus on the ResNet-18 model trained on the CIFAR-100 dataset, for which 4-bit quantization exhibits little reduction in MIA effectiveness (Table~\ref{tab:MIA-attack-4bit}). We vary the quantization bit-width from 2 to 8 bits and evaluate MIA performance across diverse quantization methods and attack techniques. Figure~\ref{fig:mia_attack_acc_cifar100_by_bits} presents the results for two representative MIAs: the black-box attack SAMIA and the white-box attack WB-P.
We observe a clear transition in the privacy impact of quantization. Specifically, 8-bit quantization has virtually no effect on MIA performance, while 4-bit quantization yields only limited and method-dependent improvements. In contrast, 2-bit quantization reduces MIA accuracy to nearly random guessing for most QNNs. These findings indicate that the privacy benefits of quantization do not increase gradually as precision decreases. Instead, meaningful reductions in membership privacy risks emerge only under extremely low precision (e.g., 2-bit quantization). However, these privacy gains come at a substantial cost in model utility, with classification accuracy dropping by as much as 95\% in some settings (as shown in Figure \ref{fig:appendix_quant_training_accuracy}).

\begin{takeaways}
\textbf{Takeaways.}
\begin{itemize}
    \item In most evaluated settings, practical quantization (4--8 bits) has little impact on membership privacy risks, with MIAs remaining nearly as effective as against full-precision models.

    \item While aggressive low-bit quantization (e.g., 2-bit) yields meaningful privacy gains, these gains come at the cost of significant utility degradation.
\end{itemize}
\end{takeaways}

\subsection{Understanding Quantization's Imact on Privacy}

To better understand why quantization has disparate effects on MIA effectiveness across different settings, we further investigate its impact on model behavior and performance. Specifically, we focus on two representative settings: 4-bit quantization of a ResNet-18 model trained on the CIFAR-100 dataset and a MobileNetV2 model trained on the CINIC-10 dataset. We consider these two settings because as shown in Table~\ref{tab:MIA-attack-4bit}, 4-bit quantization has little effect on MIA accuracy in the former setting, whereas it substantially reduces attack effectiveness in the latter.

\begin{figure}[!t]
    \centering
    \includegraphics[width=0.9\linewidth]{figures/legend_2.pdf}

    \begin{subfigure}[b]{0.42\linewidth}
        \centering
        \includegraphics[width=\linewidth]{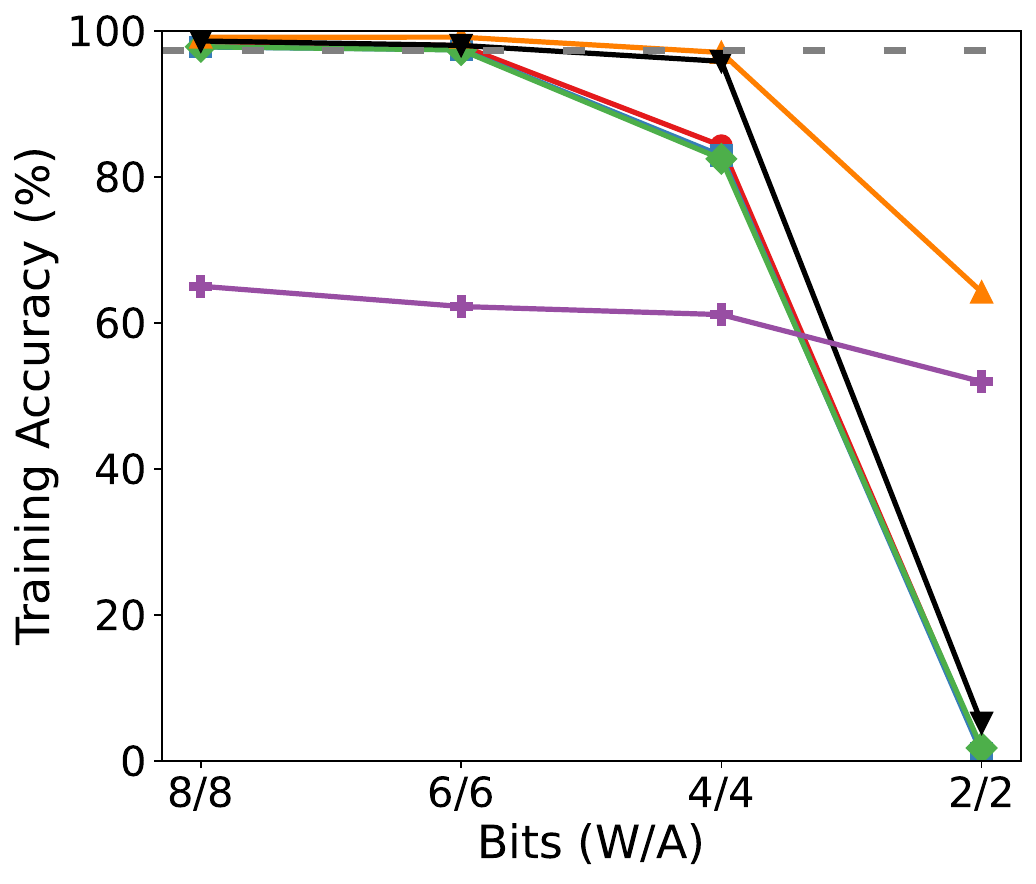}
        \caption{ResNet-18}
        \label{fig:training_acc_cifar100}
    \end{subfigure}%
    \hspace{0.2in}
    \begin{subfigure}[b]{0.42\linewidth}
        \centering
        \includegraphics[width=\linewidth]{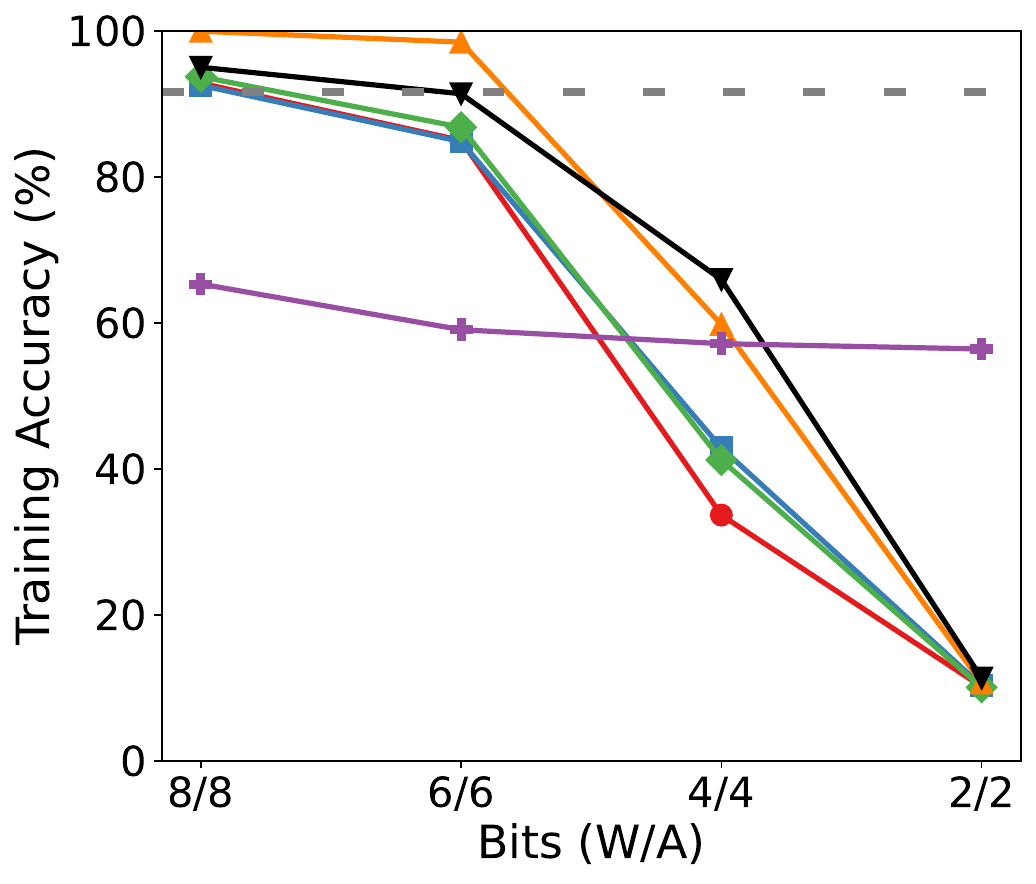}
        \caption{MobileNetV2}
        \label{fig:training_acc_cinic10_mobilenetv2}
    \end{subfigure}
    \caption{Performance of quantized models. }
    \label{fig:appendix_quant_training_accuracy}
    \vspace{-0.1in}
\end{figure}

{\bf Changes in Model Performance Do Not Explain Quantization's Privacy Effects. } The most direct effect of quantization is on the model's predictive performance. As shown in Figure~\ref{fig:appendix_quant_training_accuracy}, moderate quantization (4 bits and above) preserves training accuracy comparable to that of the corresponding full-precision models in both settings. However, more aggressive quantization introduces markedly different effects on model accuracy. At 4 bits, MobileNetV2 experiences a substantially larger accuracy drop than ResNet-18 for most quantization methods (except BSQ). When the bit-width is further reduced to 2 bits, both models suffer severe performance degradation. For example, the training accuracy of QDrop and HAWQ decreases to approximately 1\% and 10\% for ResNet-18 and MobileNetV2, respectively.
Overall, these results do not consistently correlate with its impact on MIA effectiveness. This observation suggests that the privacy effects of quantization are governed by more fundamental changes in model behavior and output rather than prediction accuracy alone.

\label{sc:effect}

\begin{figure}[!t]
    \centering
    \includegraphics[width=0.9\linewidth]{figures/legend_2.pdf}

    \begin{subfigure}[b]{0.42\linewidth}
        \centering
        \includegraphics[width=\linewidth]{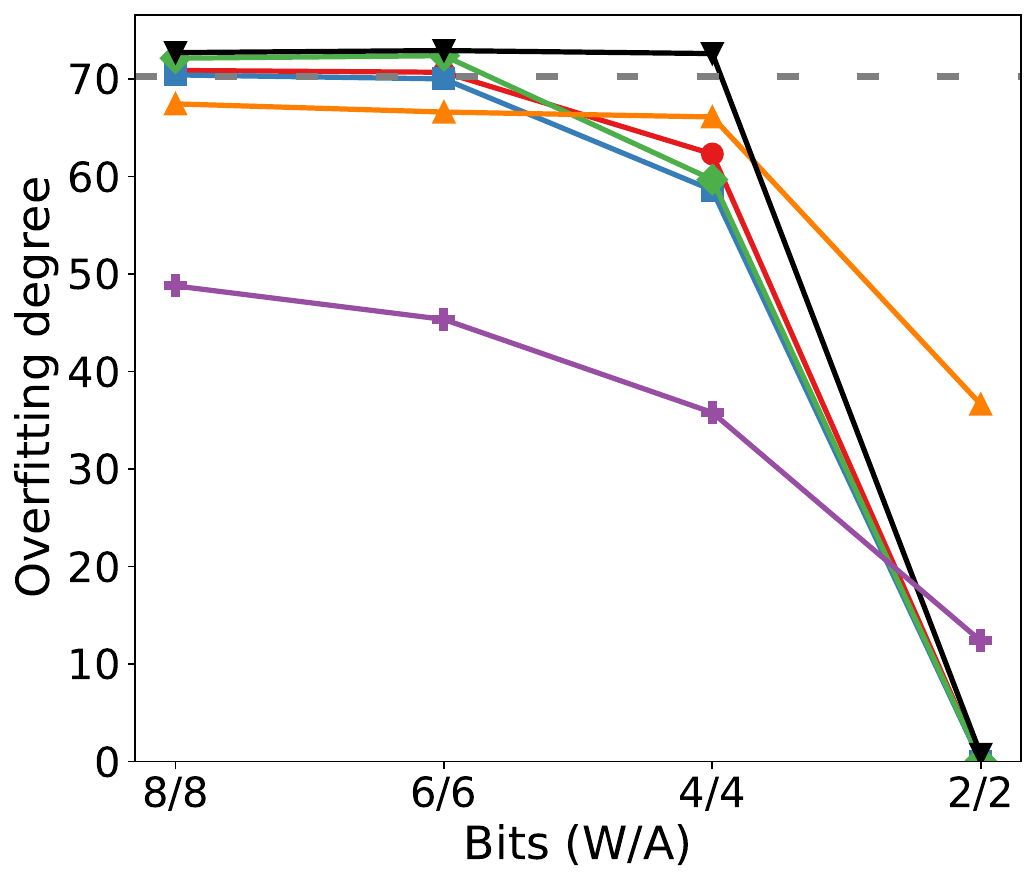}
        \caption{ResNet-18}
        \label{fig:overfitting_cifar100}
    \end{subfigure}%
    \hspace{0.2in}
    \begin{subfigure}[b]{0.42\linewidth}
        \centering
        \includegraphics[width=\linewidth]{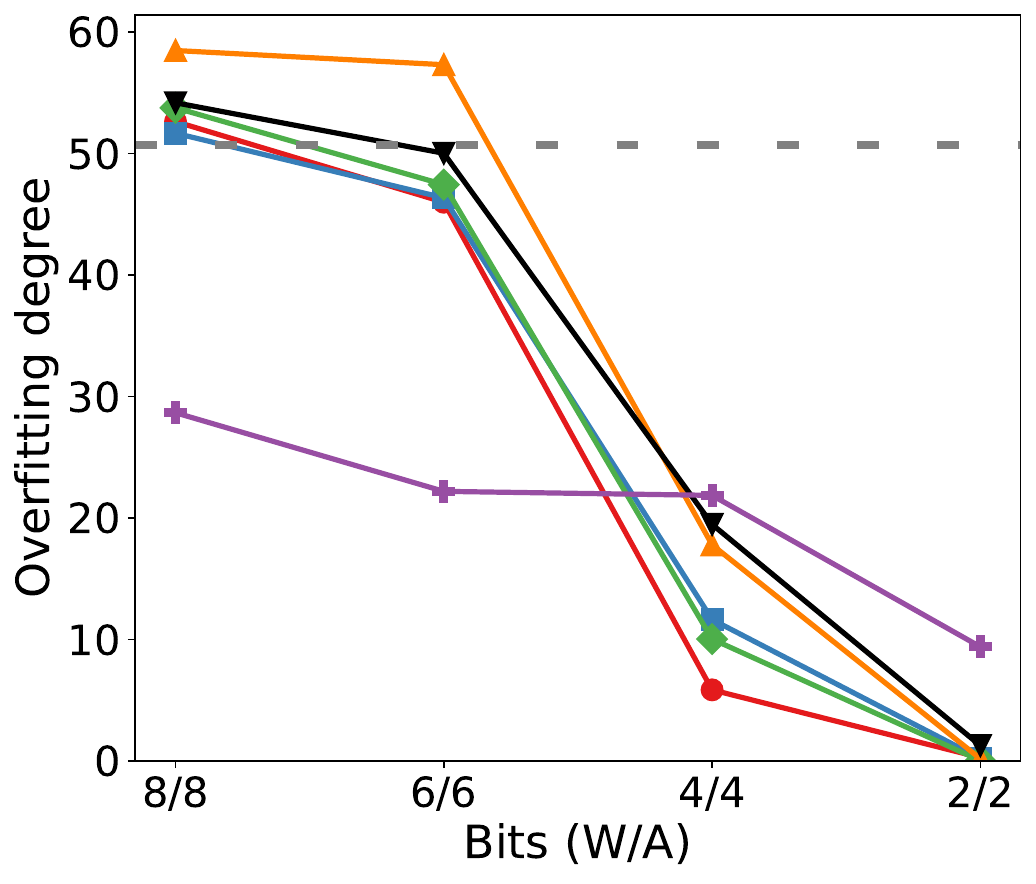}
        \caption{MobileNetV2}
        \label{fig:overfitting_cinic10}
    \end{subfigure}

    \caption{Impact of quantization on model overfitting.}
    \label{fig:model_overfitting}
    \vspace{-0.1in}
\end{figure}

\begin{figure*}[!t]
    \centering
    \captionsetup[subfigure]{font=rm,labelfont=rm,textfont=rm,skip=4pt}

    \begin{subfigure}[b]{0.26\textwidth}
        \centering
        \includegraphics[width=0.9\linewidth]{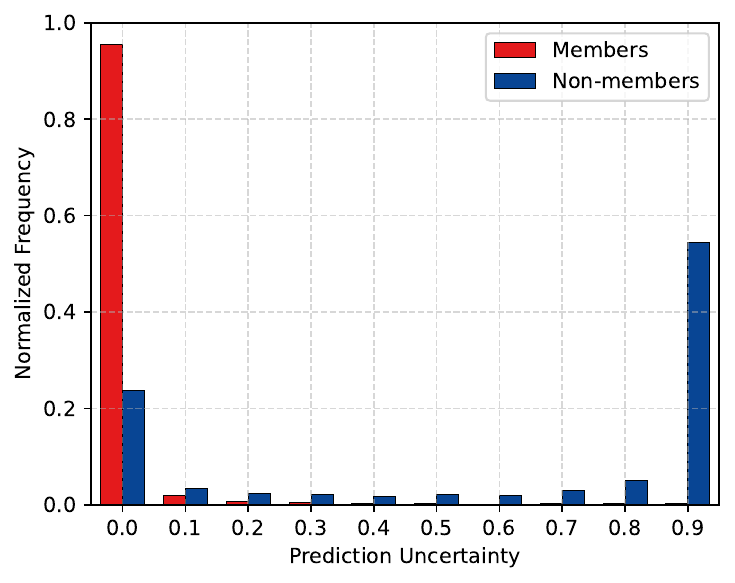}
        \caption{ResNet-18, full-precision}
        \label{fig:out_dist_cifar100_fp}
    \end{subfigure}%
    \begin{subfigure}[b]{0.26\textwidth}
        \centering
        \includegraphics[width=0.9\linewidth]{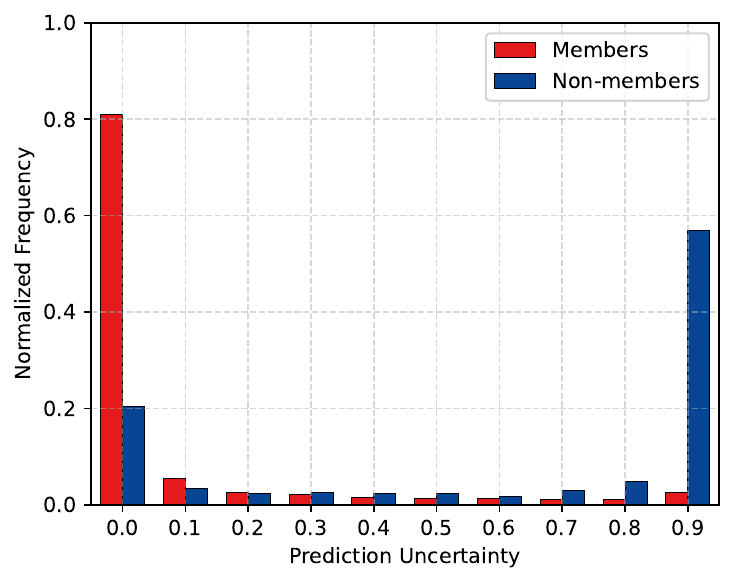}
        \caption{ResNet-18, PQ}
        \label{fig:out_dist_cifar100_pq4}
    \end{subfigure}%
    \begin{subfigure}[b]{0.26\textwidth}
        \centering
        \includegraphics[width=0.9\linewidth]{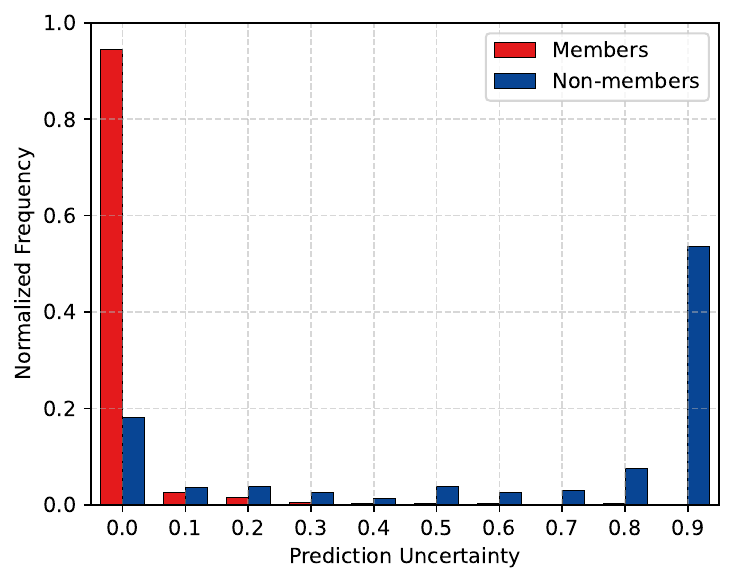}
        \caption{ResNet-18, AdaRound}
        \label{fig:out_dist_cifar100_ada4}
    \end{subfigure}

    \vspace{0.6em}

    \begin{subfigure}[b]{0.26\textwidth}
        \centering
        \includegraphics[width=0.9\linewidth]{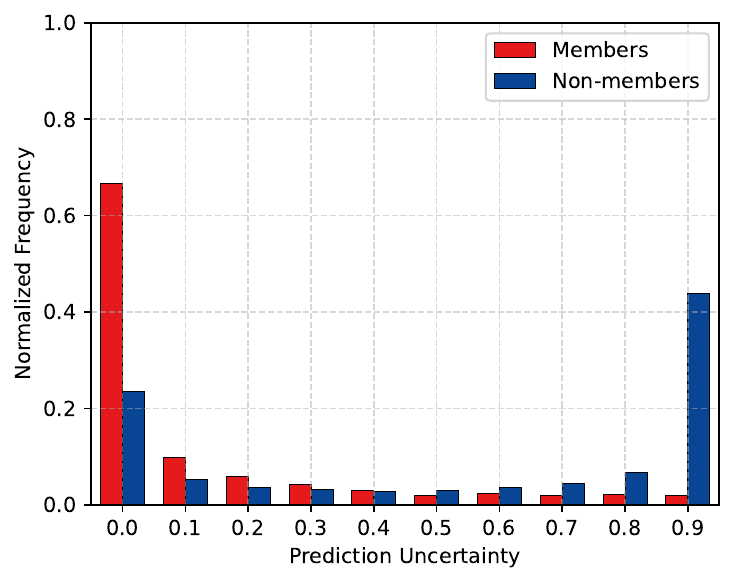}
        \caption{MobileNetV2, full-precision}
        \label{fig:out_dist_cinic10_fp}
    \end{subfigure}%
    \begin{subfigure}[b]{0.26\textwidth}
        \centering
        \includegraphics[width=0.9\linewidth]{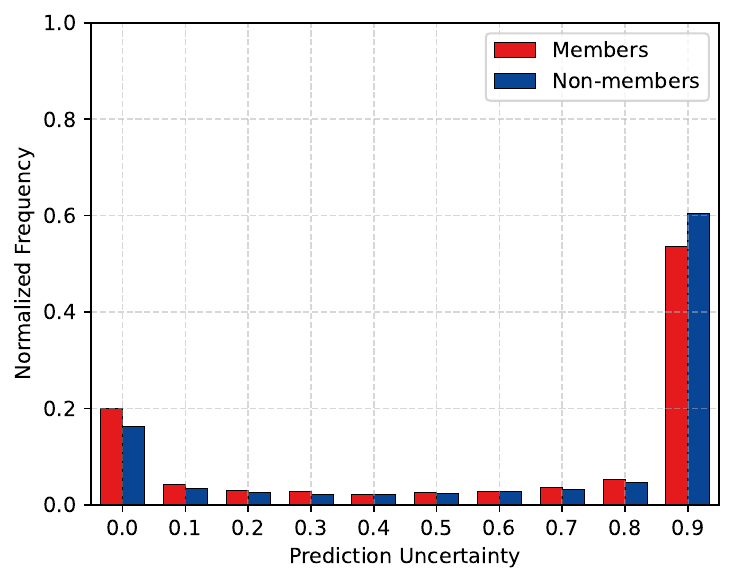}
        \caption{MobileNetV2, PQ}
        \label{fig:out_dist_cinic10_pq4}
    \end{subfigure}%
    \begin{subfigure}[b]{0.26\textwidth}
        \centering
        \includegraphics[width=0.9\linewidth]{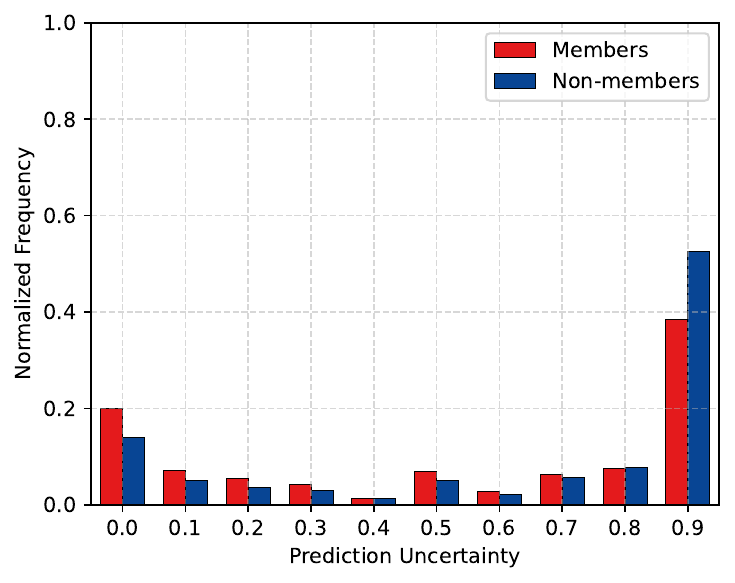}
        \caption{MobileNetV2, AdaRound}
        \label{fig:out_dist_cinic10_ada4}
    \end{subfigure}

    \caption{Output distributions (prediction uncertainty) of members and non-members before and after 4-bit quantization.}
    \label{fig:mia_output_dist_cifar100_cinic10}
\end{figure*}

{\bf Changes in Model Overfitting Only Partially Explain Quantization's Privacy Effects.} Model overfitting is measured as the gap between training and testing classification accuracy. Models with larger overfitting tend to exhibit greater membership leakage, as the behavioral differences between member and non-member samples become more distinguishable~\cite{shokri2017membership,yeom2018privacy}. Figure~\ref{fig:model_overfitting} reports the degree of overfitting under different quantization bit-widths for both ResNet-18 and MobileNetV2 models. We select these two models because 4-bit quantization exhibits contrasting effects on MIA effectiveness: it has little impact on MIA vulnerability for ResNet-18 but substantially reduces attack effectiveness for MobileNetV2 (Table~\ref{tab:MIA-attack-4bit}).
The results in Figure~\ref{fig:model_overfitting} demonstrate a clear relationship between quantization level and model overfitting. For both models, moderate quantization (6 bits and higher) preserves nearly the same degree of overfitting as full-precision models, suggesting that such compression does not substantially alter the model's effective capacity. However, reducing the bit-width to 4 bits leads to different behaviors across the two models: quantization significantly reduces overfitting for MobileNetV2, while the overfitting of ResNet-18 remains relatively unchanged. At 2-bit precision, overfitting is almost completely eliminated for both models across most quantization methods, with PACT being the only exception. These findings suggest that meaningful changes in model overfitting emerge only under aggressive quantization, and the extent of these changes depends on both the quantization level and the underlying model architecture. Such changes in overfitting can partially explain the disparate privacy effects of quantization observed across different settings.   

{\bf Changes in Output Distributions Fully Explain Quantization's Privacy Effects.} Beyond model overfitting, the output behavior of quantized models provides another important perspective for understanding their privacy implications. Since black-box MIAs primarily rely on differences in prediction confidence or uncertainty between member and non-member samples, we further analyze how quantization affects the output distributions of these two groups.
Figure~\ref{fig:mia_output_dist_cifar100_cinic10} compares the prediction-uncertainty distributions of member and non-member samples before and after 4-bit quantization for the ResNet-18 and MobileNetV2 settings. The results provide a direct explanation for the contrasting attack performance observed in these two settings (Table \ref{tab:MIA-attack-4bit}). On ResNet-18, quantization causes only minor changes to the uncertainty distributions, and the distributions of member and non-member samples remain well separated. Consequently, the prediction-confidence gap exploited by MIAs is largely preserved, allowing the attacks to remain effective, as shown in Table \ref{tab:MIA-attack-4bit}. In contrast, on MobileNetV2, the quantization substantially shifts the two distributions toward each other,  thereby weakening the membership signals and leading to a marked reduction in attack effectiveness as shown in Table \ref{tab:MIA-attack-4bit}. These observations suggest that the privacy impact of quantization is governed not merely by changes in prediction accuracy, but by how quantization reshapes the output distributions of member and non-member samples. 

\begin{takeaways}
\textbf{Takeaways.}
\begin{itemize}

    \item Reduced attack success cannot be attributed solely to degraded predictive accuracy.

    \item Although lower-precision models generally exhibit reduced overfitting, the relationship is inconsistent across model architectures and quantization methods.

    \item Quantization systematically reshapes posterior distributions, making output distributions a reliable predictor of membership inference vulnerability.
\end{itemize}
\end{takeaways}
\section{Defense Effectiveness}
\label{sec-defense}

Thus far, we have shown that QNNs under practical settings remain vulnerable to MIAs. Next, we investigate whether existing defense mechanisms that are effective for full-precision DNNs can similarly mitigate privacy risks in QNNs. 
The intuition behind this study is that quantization alters model parameters, output distributions, and internal representations, which may either strengthen or weaken the effectiveness of existing defenses. Therefore, defenses that are effective for full-precision models cannot be assumed to provide the same level of protection for QNNs. Hence, it is important to evaluate whether existing defenses remain effective after quantization.

\subsection{Defense Mechanisms}

We consider three defense mechanisms that have been widely used to protect full-precision DNNs against MIAs: 
\begin{ditemize}
\item \textbf{Dropout (Drop)}~\cite{kaya2020effectiveness,yin2021defending} randomly deactivates a percentage of neurons during each training iteration. Following the prior works ~\cite{kaya2020effectiveness,chen2024method}, we set the dropout rate $p$ to \{15\%, 20\%\}. 

\item {\bf Adversarial regularization (AR)}~\cite{nasr2018machine}
incorporates an adversarial regularizer that penalizes membership leakage into the training objective to reduce the model’s susceptibility to membership inference. 

\item {\bf Differential Privacy (DP)}~\cite{dwork2006differential}
 enforces privacy during training via DP-SGD~\cite{abadi2016deep}, which combines gradient clipping with noise injection to limit the influence of individual training samples. We consider the privacy budget $\epsilon \in \{1, 5\}$ in our evaluation. For PTQ methods (e.g., AdaRound and PQ), we first train a full-precision model with DP-SGD and subsequently apply quantization to the DP-trained weights. For QAT methods (e.g., HAWQ), we initialize from the DP-trained full-precision model and compute quantization-specific information, such as layer-wise Hessian sensitivity, using only the DP-protected model. We then jointly perform quantization-aware training and DP-SGD to maintain differential privacy throughout the training process.
\end{ditemize}

\begin{table*}[!t]
    \centering
    \small
    \caption{MIA attack accuracy (\%) under various defense mechanisms (ResNet-18, CIFAR-100).}
    \label{tab:confidence-defense}
    \resizebox{0.9\textwidth}{!}{
    \begin{tabular}{c|c|c|c|c|c|c|c|c|c|c|c}
    \hline
    \multirow{2}{*}{\textbf{Quantization}} & \multirow{2}{*}{\textbf{Bits}} & \multicolumn{5}{c|}{\textbf{C-MIA attack}} & \multicolumn{5}{c}{\textbf{WB-S attack}} \\
    \cline{3-12}
     & & \textbf{AR} & \textbf{Drop$_{20\%}$} & \textbf{Drop$_{15\%}$} & \textbf{DP$_{\epsilon=5}$} & \textbf{DP$_{\epsilon=1}$} & \textbf{AR} & \textbf{Drop$_{20\%}$} & \textbf{Drop$_{15\%}$} & \textbf{DP$_{\epsilon=5}$} & \textbf{DP$_{\epsilon=1}$} \\
    \hline
    Full-precision & 32 & 53.25 & 51.19 & 55.55 & 51.16 & 50.31 & 52.25 & 50.63 & 53.15 & 50.35 & 50.11 \\
    \hline
    \multirow{4}{*}{PQ}
    & 2 & 50.25 & 50.06 & 50.24 & 50.11 & 49.99 & 49.76 & 50.00 & 50.00 & 49.70 & 50.00 \\
    & 4 & 51.20 & 50.53 & 52.84 & 50.06 & 50.13 & 51.66 & 50.11 & 51.82 & 49.98 & 50.13 \\
    & 6 & 51.96 & 50.57 & 53.02 & 50.14 & 50.00 & 51.69 & 50.39 & 52.07 & 50.13 & 50.08 \\
    & 8 & 53.15 & 50.95 & 54.12 & 50.85 & 50.13 & 52.11 & 50.43 & 52.74 & 50.09 & 50.06 \\
    \hline
    \multirow{4}{*}{AdaRound}
    & 2 & 49.78 & 50.36 & 49.82 & 49.42 & 49.52 & 50.62 & 50.00 & 49.98 & 50.05 & 50.00 \\
    & 4 & 51.65 & 50.43 & 52.96 & 49.85 & 50.01 & 51.23 & 50.02 & 50.00 & 50.08 & 50.00 \\
    & 6 & 52.07 & 50.38 & 53.78 & 50.00 & 50.04 & 52.08 & 50.10 & 52.68 & 50.06 & 50.00 \\
    & 8 & 53.12 & 50.53 & 53.61 & 50.03 & 50.09 & 52.50 & 50.16 & 52.78 & 50.12 & 50.02 \\
    \hline
    \multirow{4}{*}{HAWQ}
    & 2 & 50.42 & 50.17 & 50.35 & 49.64 & 50.36 & 51.07 & 50.00 & 50.06 & 50.00 & 49.99 \\
    & 4 & 51.27 & 50.35 & 52.37 & 50.05 & 50.01 & 51.98 & 50.13 & 51.65 & 50.00 & 50.06 \\
    & 6 & 52.13 & 50.53 & 54.16 & 50.26 & 50.09 & 52.21 & 50.35 & 52.11 & 50.03 & 50.12 \\
    & 8 & 53.17 & 50.65 & 55.28 & 50.69 & 50.16 & 52.19 & 50.59 & 53.34 & 50.19 & 50.09 \\
    \hline
    \end{tabular}
    }
    \end{table*}

\subsection{Impact of Quantization on Defense}

{\bf Defense Effectiveness.}
Table~\ref{tab:confidence-defense} reports the MIA attack accuracy of different defense mechanisms under various quantization levels. Overall, we observe that quantization provides marginal additional privacy protection when combined with existing defense mechanisms. Compared with the corresponding full-precision models, quantized models consistently achieve slightly lower attack accuracy across most defense methods and attacks. However, the improvement is relatively limited under moderate quantization settings. For example, 8-bit quantized models exhibit attack performance close to that of full-precision models, indicating that such compression does not substantially enhance the effectiveness of existing MIA defenses.

We further observe that the privacy enhancement introduced by quantization becomes more pronounced as the quantization bit-width decreases. For all evaluated quantization methods, reducing the precision from 8 bits to 4 bits and further to 2 bits leads to a gradual reduction in MIA attack accuracy. %
This trend is consistent across both black-box (C-MIA) and white-box (WB-S) attacks, although the magnitude of improvement varies across defense mechanisms.

The observed behavior can be explained by the complementary effects of quantization and privacy defenses. All the examined MIA defenses primarily reduce membership leakage by limiting model overfitting, decreasing memorization, or perturbing model outputs. Quantization introduces an additional source of perturbation by constraining parameter precision and altering model representations. Under moderate quantization, the induced perturbation is insufficient to substantially change membership-related behaviors, and therefore provides only limited additional protection beyond the defense mechanism itself. In contrast, low-bit quantization introduces stronger parameter perturbations and significantly reduces model capacity, which can further suppress overfitting and weaken the membership signals exploited by MIAs. Consequently, the combination of aggressive quantization and existing defenses produces stronger privacy protection.

\begin{figure}[!t]
    \centering
    \includegraphics[width=1.0\linewidth]{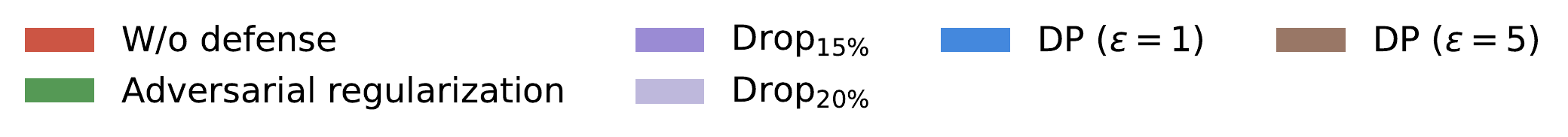}
    \captionsetup[subfigure]{font=rm,labelfont=rm,textfont=rm}
    \begin{subfigure}[b]{0.6\linewidth}
        \includegraphics[width=\linewidth]{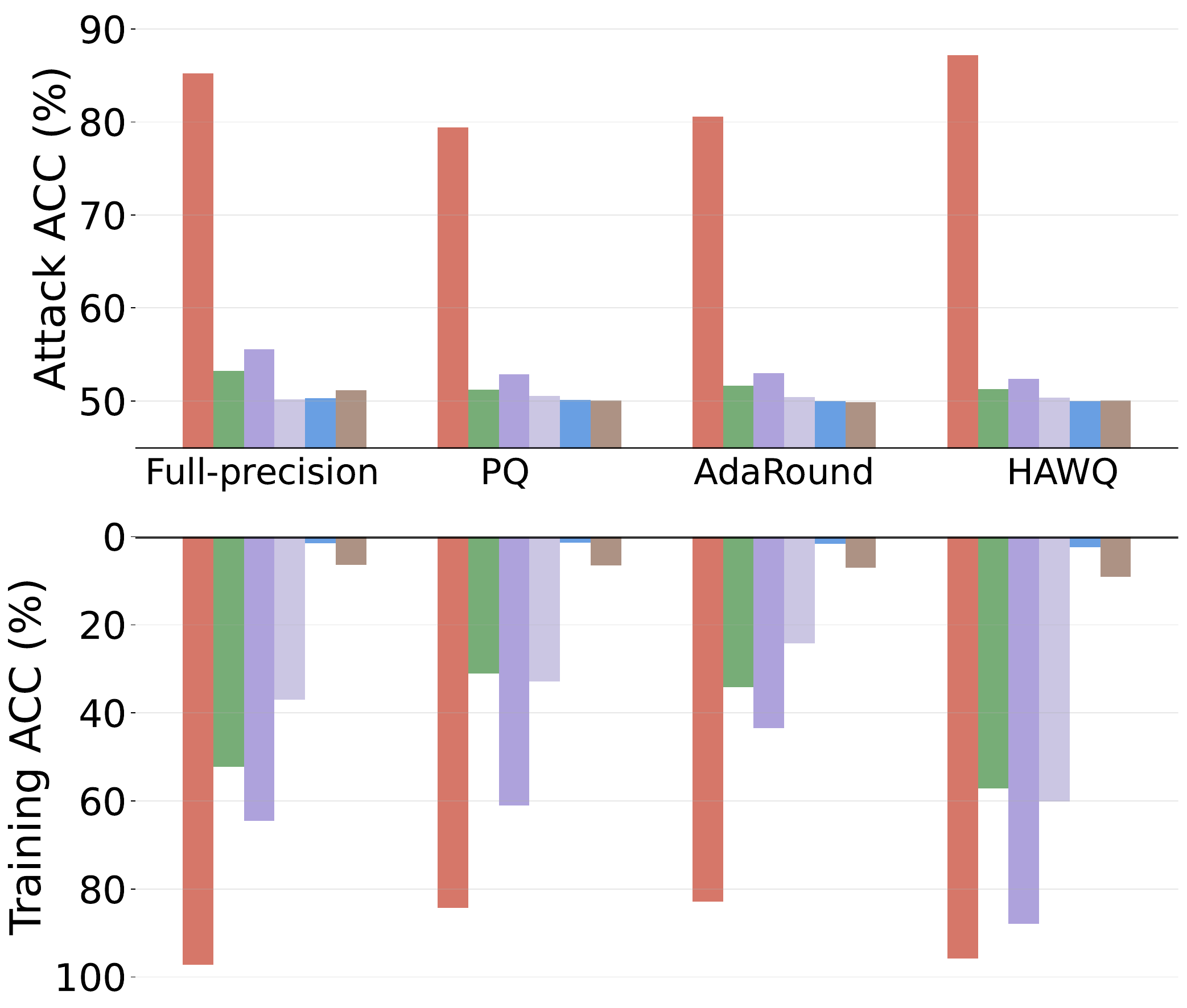}
        \label{fig:defense_conf_4bit}
    \end{subfigure}

    \caption{Trade-off between  defense effectiveness and model utility (CIFAR100, 4-bit quantization).} 
    \label{fig:defense_confidence_cifar100}
    \end{figure}

{\bf Privacy-utility Trade-off.} 
As quantization introduces additional perturbations to model parameters and representations, it remains unclear whether existing MIA defenses designed for full-precision models can achieve a favorable privacy-utility trade-off when applied to QNNs. In particular, directly combining quantization with existing defenses may amplify model distortion and further degrade predictive performance. Therefore, we evaluate the effectiveness of representative MIA defenses under quantized settings and examine whether they can provide additional privacy protection without incurring excessive utility loss.

Figure~\ref{fig:defense_confidence_cifar100} presents the privacy-utility trade-off of different defenses on full-precision and 4-bit quantized ResNet-18 models trained on the CIFAR-100 dataset. Overall, none of the evaluated defenses achieves a satisfactory balance between privacy protection and model utility. Directly applying defenses originally designed for full-precision models can further degrade the utility of QNNs due to the compounded perturbations introduced by both quantization and defense mechanisms. While DP provides the most consistent privacy protection across different settings, it incurs the most significant degradation in training accuracy. In contrast, AR and Dropout preserve higher utility, although their effectiveness varies depending on the quantization method and defense strength. Among all evaluated mechanisms, Drop$_{15\%}$ achieves the most favorable balance between privacy and utility, retaining the highest training accuracy while maintaining attack accuracy only slightly higher than the stronger privacy-preserving approaches. These results suggest that lightweight dropout defense may be more suitable for QNNs than stronger perturbation-based approaches such as DP and AR, as it  provides moderate privacy protection while avoiding excessive utility degradation.

\begin{takeaways}
\textbf{Takeaways.}
\begin{itemize}
    \item Existing MIA defenses designed for full-precision models do not directly transfer well to QNNs. They provide only limited additional privacy protection while exacerbating the utility degradation.

    \item Among the evaluated defenses, Dropout provides the best privacy-utility balance.
\end{itemize}
\end{takeaways}

\section{Future Research Directions in Building MIA-resilient QNNs}

Here, we briefly discuss several high level challenges
and opportunities for future research in building QNNs that are robust against MIAs. 

{\bf Designing New MIA Attacks against QNNs.} Our analysis reveals that quantization can reshape the privacy landscape by altering  output distributions of member and non-member samples in certain settings (Figure \ref{fig:mia_output_dist_cifar100_cinic10}). This observation suggests new opportunities for designing MIAs that explicitly exploit quantization-induced behavioral shifts. For example, the attacks could model how different quantization strategies reshape membership signals and adapt their attack features accordingly. Attackers could incorporate output calibration, distribution-shift modeling, or layer-wise sensitivity analysis to identify membership-dependent patterns introduced by low-precision representations. Some recent work \cite{li2025compleak} has explored this direction by leveraging the predictions (e.g., confidence vector) output by both full-precision and compressed models for membership inference. 
However, a fundamental challenge arises in settings where quantization has little impact on model behaviors, such as 6-bit quantization on ResNet-18 as observed in our empirical study (Figure \ref{fig:mia_attack_acc_cifar100_by_bits}). In these cases, the traditional signals exploited by MIAs remain largely preserved, leaving limited additional information for quantization-aware attacks to leverage. Developing effective MIAs in these scenarios requires exploring alternative attack surfaces beyond output distributions, such as quantization-induced parameter perturbations, representation-space changes, gradient sensitivity, or interactions between quantization artifacts and membership-dependent model responses. 

{\bf Designing Quantization-friendly Defense Mechanisms against MIAs.} Our study reveals that quantization alone is generally insufficient to provide reliable protection against MIAs under practical deployment settings.  This observation motivates the design of quantization-friendly defense mechanisms that jointly consider privacy protection and model compression. 
Our empirical studies also reveal that directly applying these defenses to QNNs introduces new challenges because quantization itself imposes additional perturbations on model parameters and representations. As shown in our previous analysis, extreme low-bit quantization can substantially reduce model accuracy by limiting the representational capacity of QNNs. 
These quantization-induced distortions can accumulate with defense-induced perturbations, potentially leading to substantial degradation in model utility. Therefore, an important research direction is to develop new privacy-preserving techniques that can effectively leverage quantization characteristics while minimizing the additional utility loss.

\section{Conclusion}\label{sec-conclusion}

In this paper, we present \system, a comprehensive framework for evaluating privacy vulnerabilities of quantized DNNs against MIAs. 
Our results show that quantization offers limited protection and can even amplify leakage, while existing defenses for full-precision models fail to consistently achieve a good privacy–utility trade-off. 
These findings highlight the need to reconsider privacy protection in quantized models and motivate a deeper understanding of the trade-offs among privacy, model accuracy, and training efficiency.

\bibliographystyle{IEEEtran}

\begin{thebibliography}{10}
\providecommand{\url}[1]{#1}
\csname url@samestyle\endcsname
\providecommand{\newblock}{\relax}
\providecommand{\bibinfo}[2]{#2}
\providecommand{\BIBentrySTDinterwordspacing}{\spaceskip=0pt\relax}
\providecommand{\BIBentryALTinterwordstretchfactor}{4}
\providecommand{\BIBentryALTinterwordspacing}{\spaceskip=\fontdimen2\font plus
\BIBentryALTinterwordstretchfactor\fontdimen3\font minus \fontdimen4\font\relax}
\providecommand{\BIBforeignlanguage}[2]{{%
\expandafter\ifx\csname l@#1\endcsname\relax
\typeout{** WARNING: IEEEtran.bst: No hyphenation pattern has been}%
\typeout{** loaded for the language `#1'. Using the pattern for}%
\typeout{** the default language instead.}%
\else
\language=\csname l@#1\endcsname
\fi
#2}}
\providecommand{\BIBdecl}{\relax}
\BIBdecl

\bibitem{rokh2023comprehensive}
B.~Rokh, A.~Azarpeyvand, and A.~Khanteymoori, ``A comprehensive survey on model quantization for deep neural networks in image classification,'' \emph{ACM Transactions on Intelligent Systems and Technology}, vol.~14, no.~6, 2023.

\bibitem{gholami2022survey}
A.~Gholami, S.~Kim, Z.~Dong, Z.~Yao, M.~W. Mahoney, and K.~Keutzer, ``A survey of quantization methods for efficient neural network inference,'' in \emph{Low-power computer vision}, 2022.

\bibitem{aubinais2025membership}
E.~Aubinais, P.~Formont, P.~Piantanida, and E.~Gassiat, ``Membership inference risks in quantized models: A theoretical and empirical study,'' \emph{arXiv preprint arXiv:2502.06567}, 2025.

\bibitem{shokri2017membership}
R.~Shokri, M.~Stronati, C.~Song, and V.~Shmatikov, ``Membership inference attacks against machine learning models,'' in \emph{2017 IEEE symposium on security and privacy (SP)}, 2017.

\bibitem{salem2019ml}
A.~Salem, Y.~Zhang, M.~Humbert, P.~Berrang, M.~Fritz, and M.~Backes, ``Ml-leaks: Model and data independent membership inference attacks and defenses on machine learning models,'' in \emph{Proceedings 2019 Network and Distributed System Security Symposium}, 2019.

\bibitem{nasr2018machine}
M.~Nasr, R.~Shokri, and A.~Houmansadr, ``Machine learning with membership privacy using adversarial regularization,'' in \emph{Proceedings of the ACM conference on computer and communications security}, 2018.

\bibitem{duddu2022inferring}
V.~Duddu and A.~Boutet, ``Inferring sensitive attributes from model explanations,'' in \emph{Proceedings of the 31st ACM International Conference on Information \& Knowledge Management}, 2022.

\bibitem{he2019model}
Z.~He, T.~Zhang, and R.~B. Lee, ``Model inversion attacks against collaborative inference,'' in \emph{Proceedings of the 35th Annual Computer Security Applications Conference}, 2019.

\bibitem{song2021systematic}
L.~Song and P.~Mittal, ``Systematic evaluation of privacy risks of machine learning models,'' in \emph{USENIX Security Symposium}, 2021.

\bibitem{carlini2022membership}
N.~Carlini, S.~Chien, M.~Nasr, S.~Song, A.~Terzis, and F.~Tramer, ``Membership inference attacks from first principles,'' in \emph{IEEE symposium on security and privacy}, 2022.

\bibitem{kowalski2022towards}
C.~Kowalski, A.~Famili, and Y.~Lao, ``Towards model quantization on the resilience against membership inference attacks,'' in \emph{IEEE International Conference on Image Processing}, 2022.

\bibitem{li2025compleak}
N.~Li, Y.~Gao, H.~Hu, B.~Kuang, and A.~Fu, ``{CompLeak}: Deep learning model compression exacerbates privacy leakage,'' in \emph{35th USENIX Security Symposium}, 2026.

\bibitem{hong2024decoding}
J.~Hong, J.~Duan, C.~Zhang, Z.~Li, C.~Xie, K.~Lieberman, J.~Diffenderfer, B.~Bartoldson, A.~Jaiswal, K.~Xu \emph{et~al.}, ``Decoding compressed trust: Scrutinizing the trustworthiness of efficient llms under compression,'' \emph{arXiv preprint arXiv:2403.15447}, 2024.

\bibitem{xu2024beyond}
Z.~Xu, A.~Gupta, T.~Li, O.~Bentham, and V.~Srikumar, ``Beyond perplexity: Multi-dimensional safety evaluation of llm compression,'' in \emph{Findings of the Association for Computational Linguistics: EMNLP 2024}, 2024, pp. 15\,359--15\,396.

\bibitem{fu2025quantized}
Y.~Fu, X.~Long, R.~Li, H.~Yu, M.~Sheng, X.~Han, Y.~Yin, and P.~Li, ``Quantized but deceptive? a multi-dimensional truthfulness evaluation of quantized llms,'' in \emph{Proceedings of the 2025 Conference on Empirical Methods in Natural Language Processing}, 2025, pp. 30\,423--30\,446.

\bibitem{famili2023deep}
A.~Famili and Y.~Lao, ``Deep neural network quantization framework for effective defense against membership inference attacks,'' \emph{Sensors}, vol.~23, no.~18, 2023.

\bibitem{jin2023membership}
Y.~Jin, W.~Lou, and Y.~Gao, ``Membership inference attacks against compression models,'' \emph{Computing. Archives for Informatics and Numerical Computation}, vol. 105, no.~11, pp. 2419--2442, 2023.

\bibitem{zhang2025bits}
C.~Zhang, T.~Qu, Z.~Li, T.~Zhang, J.~Pang, and S.~Mauw, ``Bits for privacy: Evaluating post-training quantization via membership inference,'' in \emph{2025 IEEE 24th International Conference on Trust, Security and Privacy in Computing and Communications (TrustCom)}.\hskip 1em plus 0.5em minus 0.4em\relax IEEE, 2025.

\bibitem{dwork2006differential}
C.~Dwork, ``Differential privacy,'' in \emph{International colloquium on automata, languages, and programming}, 2006.

\bibitem{abadi2016deep}
M.~Abadi, A.~Chu, I.~Goodfellow, H.~B. McMahan, I.~Mironov, K.~Talwar, and L.~Zhang, ``Deep learning with differential privacy,'' in \emph{Proceedings of the 2016 ACM SIGSAC conference on computer and communications security}, 2016.

\bibitem{kaya2020effectiveness}
Y.~Kaya, S.~Hong, and T.~Dumitras, ``On the effectiveness of regularization against membership inference attacks,'' \emph{arXiv preprint arXiv:2006.05336}, 2020.

\bibitem{yin2021defending}
Y.~Yin, K.~Chen, L.~Shou, and G.~Chen, ``Defending privacy against more knowledgeable membership inference attackers,'' in \emph{Proceedings of the 27th ACM SIGKDD Conference on Knowledge Discovery \& Data Mining}, 2021.

\bibitem{li2021deepobfuscator}
A.~Li, J.~Guo, H.~Yang, F.~D. Salim, and Y.~Chen, ``Deepobfuscator: Obfuscating intermediate representations with privacy-preserving adversarial learning on smartphones,'' in \emph{Proceedings of the International Conference on Internet-of-Things Design and Implementation}, 2021.

\bibitem{wang2021improving}
T.~Wang, Y.~Zhang, and R.~Jia, ``Improving robustness to model inversion attacks via mutual information regularization,'' in \emph{Proceedings of the AAAI Conference on Artificial Intelligence}, vol.~35, 2021.

\bibitem{hayes2024inexact}
J.~Hayes, I.~Shumailov, E.~Triantafillou, A.~Khalifa, and N.~Papernot, ``Inexact unlearning needs more careful evaluations to avoid a false sense of privacy,'' \emph{arXiv preprint arXiv:2403.01218}, 2024.

\bibitem{zarifzadeh2024low}
S.~Zarifzadeh, P.~Liu, and R.~Shokri, ``Low-cost high-power membership inference attacks,'' in \emph{International Conference on Machine Learning}, 2024.

\bibitem{leino2020stolen}
K.~Leino and M.~Fredrikson, ``Stolen memories: Leveraging model memorization for calibrated $\{$White-Box$\}$ membership inference,'' in \emph{USENIX security symposium}, 2020.

\bibitem{sablayrolles2019white}
A.~Sablayrolles, M.~Douze, C.~Schmid, Y.~Ollivier, and H.~J{\'e}gou, ``White-box vs black-box: Bayes optimal strategies for membership inference,'' in \emph{International Conference on Machine Learning}, 2019.

\bibitem{nasr2019comprehensive}
M.~Nasr, R.~Shokri, and A.~Houmansadr, ``Comprehensive privacy analysis of deep learning: Passive and active white-box inference attacks against centralized and federated learning,'' in \emph{IEEE Symposium on Security and Privacy (S\&P)}, 2019.

\bibitem{zhou2016dorefa}
S.~Zhou, Y.~Wu, Z.~Ni, X.~Zhou, H.~Wen, and Y.~Zou, ``Dorefa-net: Training low bitwidth convolutional neural networks with low bitwidth gradients,'' \emph{arXiv preprint arXiv:1606.06160}, 2016.

\bibitem{krizhevsky2009learning}
A.~Krizhevsky, G.~Hinton \emph{et~al.}, ``Learning multiple layers of features from tiny images,'' 2009.

\bibitem{darlow2018cinic}
L.~N. Darlow, E.~J. Crowley, A.~Antoniou, and A.~J. Storkey, ``Cinic-10 is not imagenet or cifar-10,'' \emph{arXiv preprint arXiv:1810.03505}, 2018.

\bibitem{he2016deep}
K.~He, X.~Zhang, S.~Ren, and J.~Sun, ``Deep residual learning for image recognition,'' in \emph{Proceedings of the IEEE conference on computer vision and pattern recognition}, 2016.

\bibitem{sandler2018mobilenetv2}
M.~Sandler, A.~Howard, M.~Zhu, A.~Zhmoginov, and L.-C. Chen, ``Mobilenetv2: Inverted residuals and linear bottlenecks,'' in \emph{Proceedings of the IEEE Conference on Computer Vision and Pattern Recognition}, 2018, pp. 4510--4520.

\bibitem{torchao}
\BIBentryALTinterwordspacing
TorchAO, ``Torchao: Pytorch-native training-to-serving model optimization,'' oct 2024. [Online]. Available: \url{https://github.com/pytorch/torchao}
\BIBentrySTDinterwordspacing

\bibitem{nagel2020up}
M.~Nagel, R.~A. Amjad, M.~Van~Baalen, C.~Louizos, and T.~Blankevoort, ``Up or down? adaptive rounding for post-training quantization,'' in \emph{International conference on machine learning}, 2020.

\bibitem{wei2022qdrop}
X.~Wei, R.~Gong, Y.~Li, X.~Liu, and F.~Yu, ``Qdrop: Randomly dropping quantization for extremely low-bit post-training quantization,'' \emph{arXiv preprint arXiv:2203.05740}, 2022.

\bibitem{choi2018pact}
J.~Choi, Z.~Wang, S.~Venkataramani, P.~I.-J. Chuang, V.~Srinivasan, and K.~Gopalakrishnan, ``Pact: Parameterized clipping activation for quantized neural networks,'' \emph{arXiv preprint arXiv:1805.06085}, 2018.

\bibitem{yang2021bsq}
H.~Yang, L.~Duan, Y.~Chen, and H.~Li, ``Bsq: Exploring bit-level sparsity for mixed-precision neural network quantization,'' \emph{arXiv preprint arXiv:2102.10462}, 2021.

\bibitem{dong2019hawq}
Z.~Dong, Z.~Yao, A.~Gholami, M.~W. Mahoney, and K.~Keutzer, ``Hawq: Hessian aware quantization of neural networks with mixed-precision,'' in \emph{Proceedings of the IEEE/CVF international conference on computer vision}, 2019.

\bibitem{yuan2022mia_pruning}
X.~Yuan and L.~Zhang, ``Membership inference attacks and defenses in neural network pruning,'' in \emph{USENIX Security Symposium}, 2022.

\bibitem{chen2021machine}
M.~Chen, Z.~Zhang, T.~Wang, M.~Backes, M.~Humbert, and Y.~Zhang, ``When machine unlearning jeopardizes privacy,'' in \emph{Proceedings of the 2021 ACM SIGSAC conference on computer and communications security}, 2021.

\bibitem{yeom2018privacy}
S.~Yeom, I.~Giacomelli, M.~Fredrikson, and S.~Jha, ``Privacy risk in machine learning: Analyzing the connection to overfitting,'' in \emph{IEEE Computer security foundations symposium (CSF)}, 2018.

\bibitem{yao2021hawq}
Z.~Yao, Z.~Dong, Z.~Zheng, A.~Gholami, J.~Yu, E.~Tan, L.~Wang, Q.~Huang, Y.~Wang, M.~Mahoney \emph{et~al.}, ``Hawq-v3: Dyadic neural network quantization,'' in \emph{International Conference on Machine Learning}.\hskip 1em plus 0.5em minus 0.4em\relax PMLR, 2021, pp. 11\,875--11\,886.

\bibitem{dong2022finding}
R.~Dong, Z.~Tan, M.~Wu, L.~Zhang, and K.~Ma, ``Finding the task-optimal low-bit sub-distribution in deep neural networks,'' in \emph{International Conference on Machine Learning}.\hskip 1em plus 0.5em minus 0.4em\relax PMLR, 2022, pp. 5343--5359.

\bibitem{chen2024method}
Z.~Chen and K.~Pattabiraman, ``A method to facilitate membership inference attacks in deep learning models,'' \emph{arXiv preprint arXiv:2407.01919}, 2024.

\end{thebibliography}

\appendices

\begin{table*}[!t]
    \centering
    \caption{TPR@1\%FPR (\%) of MIAs across 4-bit quantized models in different settings}.
    \label{tab:tpr1-4bit}
    \begin{tabular}{c|c|c|c|c|c|c|c|c|c}
    \hline
    \multirow{2}{*}{\textbf{\makecell{Dataset\&\\ Model}}} & \multirow{2}{*}{\textbf{Quantization}}
    & \multicolumn{6}{c|}{\textbf{Black-box MIA}} 
    & \multicolumn{2}{c}{\textbf{White-box MIA}} \\
    \cline{3-10}
    & & \textbf{C-MIA} & \textbf{Ent} & \textbf{M-Ent} & \textbf{ML-leaks} & \textbf{SAMIA} & \textbf{RMIA} & \textbf{WB-S} & \textbf{WB-P} \\
    \hline
    
    \multirow{7}{*}{\makecell{CIFAR-100\\ ResNet-18}}
    & FP
    & 0.74 & 0.69 & 0.77 & 2.93 & 2.18 & 51.80 & 81.42 & 82.36 \\\cline{2-10}
    & PQ
    & 0.51 & 0.45 & 0.55 & 1.34 & 1.16 & 42.60 & 68.40 & 69.18 \\
    & AdaRound
    & 0.56 & 0.50 & 0.60 & 1.48 & 1.27 & 44.30 & 69.20 & 70.05 \\
    & QDrop
    & 0.54 & 0.48 & 0.58 & 1.56 & 1.35 & 46.80 & 71.60 & 71.28 \\
    & HAWQ
    & 0.69 & 0.63 & 0.72 & 2.52 & 1.96 & 50.10 & 80.30 & 81.21 \\
    & PACT
    & 0.73 & 0.67 & 0.75 & 2.68 & 2.07 & 50.70 & 79.80 & 80.74 \\
    & BSQ
    & 0.21 & 0.18 & 0.24 & 0.76 & 0.64 & 18.40 & 34.51 & 35.26 \\
    \hline
    
    \multirow{7}{*}{\makecell{CINIC-10\\MobileNetV2}}
    & FP
    & 0.82 & 0.70 & 0.79 & 1.92 & 2.38 & 7.84 & 9.68 & 10.06 \\\cline{2-10}
    & PQ
    & 0.52 & 0.47 & 0.51 & 1.02 & 1.28 & 4.36 & 4.95 & 5.19 \\
    & AdaRound
    & 0.60 & 0.54 & 0.59 & 1.26 & 1.55 & 5.42 & 5.96 & 5.82 \\
    & QDrop
    & 0.58 & 0.52 & 0.57 & 1.21 & 1.50 & 5.18 & 5.72 & 5.99 \\
    & HAWQ
    & 0.70 & 0.62 & 0.69 & 1.58 & 1.86 & 6.48 & 6.95 & 7.28 \\
    & PACT
    & 0.72 & 0.64 & 0.71 & 1.64 & 1.93 & 6.62 & 7.08 & 7.41 \\
    & BSQ
    & 0.56 & 0.50 & 0.55 & 1.18 & 1.48 & 5.34 & 5.88 & 6.16 \\
    \hline
    
    \multirow{7}{*}{\makecell{Texas\\ MLP}}
    & FP
    & 0.78 & 0.63 & 0.75 & 1.94 & 2.24 & 10.84 & 13.42 & 14.06 \\\cline{2-10}
    & PQ
    & 0.70 & 0.57 & 0.68 & 1.67 & 1.96 & 9.75 & 12.10 & 11.92 \\
    & AdaRound
    & 0.72 & 0.59 & 0.70 & 1.74 & 2.01 & 9.92 & 12.38 & 12.91 \\
    & QDrop
    & 0.71 & 0.58 & 0.69 & 1.70 & 1.98 & 9.86 & 12.27 & 12.84 \\
    & HAWQ
    & 0.75 & 0.61 & 0.73 & 1.86 & 2.14 & 10.42 & 13.08 & 13.62 \\
    & PACT
    & 0.73 & 0.60 & 0.71 & 1.81 & 2.08 & 10.15 & 12.84 & 13.38 \\
    & BSQ
    & 0.61 & 0.49 & 0.59 & 1.42 & 1.71 & 8.54 & 10.72 & 11.16 \\
    \hline
    
    \multirow{7}{*}{\makecell{Location\\MLP}}
    & FP
    & 0.75 & 0.64 & 0.73 & 2.48 & 8.76 & 17.54 & 17.20 & 17.95 \\\cline{2-10}
    & PQ
    & 0.53 & 0.47 & 0.52 & 1.65 & 5.52 & 11.95 & 12.36 & 12.94 \\
    & AdaRound
    & 0.55 & 0.49 & 0.54 & 1.70 & 5.70 & 12.12 & 12.58 & 13.15 \\
    & QDrop
    & 0.52 & 0.46 & 0.51 & 1.60 & 5.18 & 11.52 & 11.98 & 12.52 \\
    & HAWQ
    & 0.50 & 0.45 & 0.49 & 1.58 & 5.00 & 11.18 & 11.82 & 11.54 \\
    & PACT
    & 0.69 & 0.60 & 0.68 & 2.24 & 7.55 & 13.36 & 15.42 & 15.88 \\
    & BSQ
    & 0.34 & 0.30 & 0.35 & 1.14 & 3.80 & 7.82 & 8.38 & 8.56 \\
    \hline
    \end{tabular}    
    \end{table*}

\section{Details of Datasets and DNN models}
\label{sec:appendix_dataset_stats}

\begin{table}[!t]
\centering
\caption{Statistics of Datasets}
\label{tab:dataset_stats}
\scalebox{0.95}{
\begin{tabular}{ccccc}
\toprule
\textbf{Dataset} & \textbf{Feature dim.} & \textbf{\# labels} & \textbf{\# train samples} & \textbf{\# test samples} \\
\midrule
CIFAR-100 & $32 \times 32 \times 3$ & 100 & 50,000 & 10,000 \\
CINIC-10 & $32 \times 32 \times 3$ & 10 & 90,000 & 90,000 \\ 
Location & 446 & 30 & 4,008 & 1,002 \\
Texas & 6,169 & 100 & 53,864 & 13,466 \\
\bottomrule
\end{tabular}
}
\end{table}

{\bf Datasets.} Table~\ref{tab:dataset_stats} summarizes the statistics of the datasets used in \system. 

{\bf DNNs.} \textit{ResNet-18} consists of 17 convolutional layers and one fully connected layer, organized into four residual stages with two basic blocks per stage, followed by global average pooling and a final classification layer. \textit{MobileNetV2} is a lightweight convolutional neural network built upon inverted residual blocks and linear bottlenecks. It uses depthwise separable convolutions to reduce computational cost while maintaining strong classification performance. For MLP, we use a five-layer fully connected neural network with layer sizes [2048, 1024, 512, 256, 100] and Tanh activation functions. All parameters of the MLP are initialized from a random normal distribution with mean 0 and standard deviation 0.01.

\section{Attack TPR@1\% FPR Results Results}
\label{sec:appendix_mia_acc_cinic10}
\label{sec:appendix_mia_auc}

Table~\ref{tab:tpr1-4bit} reports TPR@1\% FPR for membership inference attacks across two dataset-model settings. The trends are similar to the attack accuracy results in Section~\ref{sec-evaluation} and are therefore omitted.

\end{document}